\documentclass[preprintnumbers, amssymb,amsmath,aps,prl,twocolumn,floatfix,nofootinbib,superscriptaddress,showpacs]{revtex4-1}

\usepackage{epsfig}
\usepackage{bm}
\usepackage{amssymb}
\usepackage{amsmath}
\usepackage[dvipsnames]{xcolor}
\usepackage{subfigure}
\usepackage[
  colorlinks=true,
  linkcolor=blue,
  anchorcolor=black,
  citecolor=blue,
  urlcolor=blue
]{hyperref}
\usepackage{soul}
\usepackage[capitalise]{cleveref}
\usepackage[scr=boondoxo]{mathalfa}
\usepackage{booktabs}
\usepackage{tabularx}

\usepackage{lmodern}
\usepackage[scaled]{helvet}
\usepackage{lettrine}
\usepackage{csquotes}
\usepackage[top=0.8in, bottom=0.8in, left=0.5in, right=0.5in, columnsep=0.25in]{geometry}

\usepackage{fontawesome}
\usepackage{eso-pic,graphicx}
\usepackage{orcidlink}

\newcommand{\LogoOffsetX}{0.5in}
\newcommand{\VertOffset}{1.85cm}

\AddToShipoutPictureBG*{
  \AtPageUpperLeft{
    \raisebox{-\VertOffset}{
      \makebox[\paperwidth][l]{
        \hspace{\LogoOffsetX}\includegraphics[width=4cm]{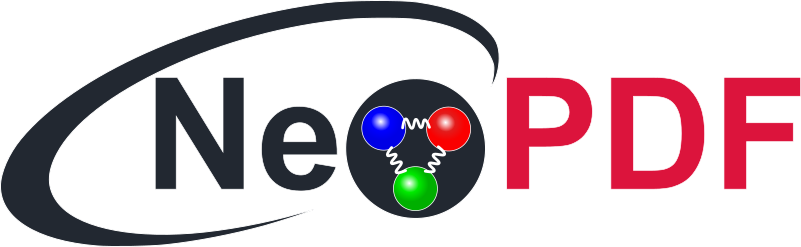}
        \hfill
        {NIKHEF-2025-014}\hspace{\LogoOffsetX}
      }
    }
  }
}

\usepackage[authormarkup=brackets,commentmarkup=uwave,deletedmarkup=sout]{changes}
\definechangesauthor[color=NavyBlue]{YH}
\setauthormarkup{ }
\definechangesauthor[color=magenta]{XB}
\definechangesauthor[color=blue]{HL}

\newcommand{\neopdf}{\textsc{NeoPDF}}
\newcommand{\lhapdf}{\textsc{LHAPDF}}
\newcommand{\tmdlib}{\textsc{TMDlib}}

\usepackage{listings}

\lstdefinelanguage{Mathematica}{
  morekeywords={
    Plot,Table,Integrate,Sum,Module,Block,Do,While,For,If,Switch,Which,
    ReplaceAll,ReplaceRepeated,RuleDelayed,Rule,
    Function,SetDelayed,Set,Clear,ClearAll,Print,Return,
    True,False,Null,Infinity,
    LibraryFunctionLoad
  },
  sensitive=true,
  morecomment=[s]{(*}{*)},
  morestring=[b]",
  alsoletter={_}
}

\definecolor{mygreen}{rgb}{0,0.6,0}
\definecolor{mygray}{rgb}{0.5,0.5,0.5}
\definecolor{mymauve}{rgb}{0.58,0,0.82}
\newcommand{\codify}[1]{\textcolor{black}{\texttt{#1}}}

\renewcommand\thesection{\arabic{section}}
\renewcommand\thesubsection{\thesection.\arabic{subsection}}
\renewcommand\thesubsubsection{\thesubsection.\arabic{subsubsection}}	

\makeatletter
\renewcommand\frontmatter@abstractfont{%
	\normalfont\normalsize
	\parindent1em\relax
	\adjust@abstractwidth
}
 \renewcommand\frontmatter@abstractwidth{.8\textwidth}

\def\@hangfrom@section#1#2{#1#2}
\def\@section@style#1{%
	\@hangfrom@section{}{%
		\normalfont\bfseries\raggedright\thesection\quad#1}%
}
\def\@seccntformat#1{\csname the#1\endcsname\quad}

\renewcommand\section{\@startsection{section}{1}{0pt}%
	{-3.5ex \@plus -1ex \@minus -.2ex}%
	{2.3ex \@plus .2ex}%
	{\normalfont\bfseries\raggedright}}

\renewcommand\subsection{\@startsection{subsection}{2}{0pt}%
	{-3.25ex \@plus -1ex \@minus -.2ex}%
	{1.5ex \@plus .2ex}%
	{\normalfont\bfseries\raggedright}}

\renewcommand\subsubsection{\@startsection{subsubsection}{3}{0pt}%
	{-3.25ex \@plus -1ex \@minus -.2ex}%
	{1.5ex \@plus .2ex}%
	{\normalfont\itshape\raggedright}}
	
\renewcommand\paragraph{\@startsection{paragraph}{4}{\z@}%
	{1.0ex \@plus 0.5ex \@minus.2ex}%
	{-1em}%
	{\normalfont\bfseries\scshape}}

\def\maketitle{ %
\@author@finish
\title@column\titleblock@produce
\suppressfloats[t]
  \let\and\relax
  \let\affiliation\@gobble
  \let\@AAC@list\@empty
  \let\@AFF@list\@empty
  \let\@AFG@list\@empty
  \let\@AF@join\@AF@join@error
  \let\@address\@empty
  \let\thanks\@gobble
  \let\abstract\@undefined\let\endabstract\@undefined
  \titlepage@sw{%
   \vfil
   \clearpage
  }{}%
}%
\makeatother

\begin{document}

\vspace*{0.5cm}

\title{\neopdf: A fast interpolation library for collinear and transverse \\ momentum-dependent parton distributions}

\author{Tanjona R. Rabemananjara~\!\orcidlink{0000-0002-8395-8059}}\email{tanjona.hepc@gmail.com}
\affiliation{Nikhef Theory Group, Science Park 105, 1098 XG Amsterdam, The Netherlands}
\affiliation{Department of Physics and Astronomy, Vrije Universiteit, NL-1081 HV Amsterdam}

\begin{abstract}
We present \neopdf, an interpolation library that supports both collinear and transverse momentum-dependent parton distribution
functions. \neopdf~is designed to be fast and reliable, with modern functionalities that target both current and future hadron collider
experiments. It aims to address the shortcomings of existing interpolation libraries while providing additional features to support
generic non-perturbative functions.
Some of the features include a new interpolation based on Chebyshev polynomials, as well as the ability to interpolate along
the nucleon number $A$, the reference strong coupling $\alpha_s(M_Z)$, and the parton's intrinsic transverse momentum $k_T$.
\neopdf~implements its own file format using binary serialisation and lossless compression, prioritising speed and efficiency.
A no-code migration design is provided for \lhapdf~in order to remove the frictions associated with transitioning to \neopdf.
The library is written in Rust with interfaces for various programming languages such as Fortran, C, C++, Python, and Mathematica.
We benchmark \neopdf~against \lhapdf~and \tmdlib~for various sets and show that it is both fast and accurate.
\end{abstract}

\maketitle


\section{Introduction}

Parton Density Functions (PDFs) constitute a central ingredient in the phenomenology of Quantum Chromodynamics (QCD), encoding the
non-perturbative structure of hadrons in terms of their partonic degrees of freedom. In light of the present work, "PDFs" shall refer to
the collection of non-perturbative functions with which partonic cross-sections must be convolved to obtain physical results that can
be compared to experimental data.

The most commonly employed functions are the unpolarised proton PDFs~\cite{Gao:2017yyd,Kovarik:2019xvh,Forte:2013wc}, which describe
the longitudinal momentum distribution of quarks and gluons (and leptons) found in an unpolarised proton. Their polarised counterparts
(pPDFs)~\cite{Aidala:2012mv,Jimenez-Delgado:2013sma,Borsa:2024mss,Cruz-Martinez:2025ahf,Bertone:2024taw} capture the spin-dependent
structure and are essential for understanding the helicity and transversity of partons.
Beyond the proton, nuclear PDFs (nPDFs)~\cite{Ethier:2020way,Klasen:2023uqj} account for the modifications of parton distributions
in bound nucleons, thereby enabling the studies of nuclear effects in high-energy scattering.
All of these quantities arise in \enquote{spacelike} factorisation where the relevant hard probe (e.g. a virtual photon for deeply
inelastic scatterings) carries negative invariant mass squared ($q^2 < 0$).
On the other hand, Fragmentation Functions (FFs)~\cite{Metz:2016swz,Bertone:2018ecm,AbdulKhalek:2022laj,Khalek:2021gxf} characterise the hadronisation
process, describing how energetic partons evolve into final-state hadrons. Analogous to the initial-state parton densities, they can
also be unpolarised or polarised. FFs are associated with \enquote{timelike} factorisation in which the hard probe (parton) carries
positive invariant mass squared ($q^2 > 0$).

From a kinematic perspective, initial- and final-state parton densities are usually formulated in the collinear approximation whereby
they are only defined in terms of the longitudinal momentum. They can be generalised to Transverse Momentum-Dependent distributions
(TMDs)~\cite{Boussarie:2023izj,Angeles-Martinez:2015sea}, which retain the information on the intrinsic transverse momentum and are crucial
for a three-dimensional mapping of hadrons in momentum space.
Analogous to the collinear case, TMDs can be defined for both the proton and nuclei, providing insights into medium modifications of parton
dynamics in nuclear environments. They can also be classified as unpolarised or polarised, encoding correlations between spin and transverse
momentum such as in the Sivers~\cite{Sivers:1989cc,Echevarria:2020hpy} or Boer–Mulders~\cite{Boer:1997nt,LatticeParton:2024mxp} functions.
Spacelike TMDs (TMDPDFs) appear in processes such as Deep Inelastic Scattering (DIS), where they describe the initial-state partonic
structure of the hadron, while timelike TMDs (TMDFFs) arise in processes such as back-to-back hadron production in $e^+e^-$ annihilation,
where they characterise the fragmentation of final-state partons. Similar to the collinear PDF case, TMDPDFs and TMDFFs can simultaneously
arise in processes such as Semi-Inclusive Deep Inelastic Scattering (SIDIS). Together, these various classes of TMDs provide a unified
framework for investigating both the spin–momentum correlations and the nuclear dependence of hadron structure.

A further generalisation of these non-perturbative functions is provided by the Generalised Transverse-Momentum Dependent distributions
(GTMDs)~\cite{Lorce:2011dv,Bhattacharya:2017bvs}, which represent the most comprehensive class of parton correlation functions. GTMDs
encode information on the longitudinal momentum, intrinsic transverse momentum, and spatial distributions of partons, thereby offering
a full five-dimensional phase-space picture of hadron structure in QCD.
In specific limits, GTMDs reduce to other, more familiar partonic distributions. In the forward limit, while retaining transverse momentum, they reduce
to TMDs. When integrated over the transverse momentum, they reduce to Generalised Parton Distributions
(GPDs)~\cite{Mezrag:2023nkp,Diehl:2003ny,Guidal:2013rya,Ji:2004gf}, which preserve spatial information.
Finally, in the forward and collinear limit, GTMDs reduce to ordinary parton density functions. In this sense, GTMDs serve as the most general
representation from which all other partonic distributions can be derived.

The wide variety of PDF classes---ranging from collinear parton densities and their polarised and nuclear extensions to (G)TMDs and FFs---captures
complementary aspects of QCD dynamics. Traditionally, each class of distributions is defined and analysed within its own framework, with limited
connections. For instance, the  \lhapdf~interpolation library~\cite{Buckley:2014ana} is only used for collinear distributions, while its counterpart
\tmdlib~\cite{Hautmann:2014kza,Abdulov:2021ivr} is mainly used for transverse momentum-dependent distributions. This, however, complicates
efforts to build a coherent picture of hadron structure across different processes and kinematic domains.
A unified interpolation framework is therefore highly desirable to enable consistent treatment of the non-perturbative functions  as interconnected
manifestations of partonic structure.

The \neopdf~library\footnote{\faGithub~\textbf{Code Repository}: \href{https://github.com/QCDLab/neopdf}{https://github.com/QCDLab/neopdf}} aims
to address this need by offering a computational infrastructure that interpolates smoothly between the various classes
of distributions. By treating the different functions as limits or projections of a generalized underlying structure,  \neopdf~provides a consistent
and flexible framework for QCD phenomenology in the precision era.
This complements and extends the efforts towards simultaneous fits, by offering a unified framework that can integrate diverse data sets and
theoretical inputs into a single, coherent description of hadron structure.

\neopdf~distinguishes itself from other interpolation libraries~\cite{Buckley:2014ana,Hautmann:2014kza,Abdulov:2021ivr,KordValeshabadi:2024bxh} in
two main aspects. The first aspect is of physics nature. On the one hand, \neopdf~is fully
compatible with LHAPDF and its format in that it can interpolate functions along the momentum fraction $x$ and scale $Q^2$. In addition,
it also supports interpolations across the nucleon number $A$ and the strong coupling $\tilde{\alpha}_s \equiv \alpha_s(M_Z)$. While the former is
crucial in the context of nuclear PDF fits, the latter is relevant in a simultaneous extraction of $\alpha_s$ and PDFs (or other QCD parameters extractions)
and for assessing theoretical uncertainties arising in perturbative computations.

On the other hand, \neopdf~supports TMD distributions by being able to additionally interpolate across the intrinsic transverse momentum $k_T$
of the parton, making it also an alternative to \tmdlib.
In the future, \neopdf~will be extended to support GTMDs (and as a consequence GPDs) by accounting for the skewness $\xi$---which quantifies the
longitudinal momentum transfer between the initial and final hadron states in an exclusive scattering process---and the total momentum transfer
$t$. However, this will be left for future work.

The second aspect is of technical nature. A distinctive feature of \neopdf~lies in its ability to clearly distinguish between the types of distributions.
This makes convolutions with partonic cross-sections unambiguous when dealing with processes that involve multiple types of initial-/final-state
hadrons by ensuring that convolution routines automatically select and apply the correct non-perturbative input, thereby minimizing user-side ambiguity.
This is particularly relevant when the partonic cross-sections are available in terms of fast interpolation
grids~\cite{Carrazza:2020gss,Carli:2010rw,Kluge:2006xs,Wobisch:2011ij,Britzger:2012bs,christopher_schwan_2025_15635174} where information about the
types of hadrons are not readily available.

Another technical aspect that distinguishes \textsc{NeoPDF} is the implementation of interpolations based on Chebyshev polynomials. Such interpolation
methods have been proposed in~\cite{Diehl:2021gvs} as an efficient method to interpolate PDFs. As we will demonstrate in the paper, the multi-dimensional
Chebyshev interpolation
routine implemented in \neopdf~easily outperforms piecewise cubic spline interpolations (such as the one used as default in \lhapdf) by orders of
magnitude in accuracy with lower number of grid points.
If specific interpolation strategies are needed, these can be seamlessly implemented within the codebase thanks to the modular architecture and
extensibility of \neopdf.

The core library of \neopdf~is written in Rust in order to leverage Rust's zero-cost abstractions, speed, and safety. The core functionalities are exposed
to the Fortran, C/C++, Python, and Mathematica Application Programming Interfaces (APIs) for a seamless interoperability with other codes that can link to these
programming languages.

The remainder of the paper is organised as follows. In Section~\ref{sec:design} we describe in details the \neopdf~design structure and rationale,
emphasizing on the new features it provides. We also discuss the \neopdf~grid data representation and the interpolations it currently supports. In
Section~\ref{sec:benchmark} we benchmark the accuracies and speed of \neopdf~against other interpolation libraries, namely \lhapdf~for collinear
PDFs and \tmdlib~for transverse momentum-dependent ones. We conclude in Section~\ref{sec:conclusions} and provide some perspectives for
future developments.
Finally, examples of usage in various programming languages and the installation of \neopdf~are provided in the Appendix.

\section{The NeoPDF design}
\label{sec:design}

The following section describes the design rationale that underpins the \neopdf~library, its features, and what sets it apart from existing
interpolation libraries. Details on the grid representation, interpolation strategies, and the \neopdf~file format are also provided.

\subsection{Design rationale}
\label{subsec:design}

The objects that \neopdf~ought to describe are non-perturbative functions---that we collectively refer to as \enquote{PDFs}---with which partonic
cross-sections are convolved to obtain physics observables. In the context of collinear and transverse momentum-dependent parton densities, 
for a given parton flavour $i$, such functions generally take the following form:
\begin{equation}
	\left( \Delta \right) f_i \left( A, x, k_T, Q^2, \tilde{\alpha}_s \right),
	\label{eq:pdfs-repr}
\end{equation}
where $A$ is the atomic mass number, $x$ is the longitudinal momentum fraction, $k_T$ the intrinsic transverse momentum of the parton, and $Q^2$
the energy scale at which the cross-sections are factorised. The (un)polarised parton density function $(\Delta)f_i$ also has an intrinsic dependence
on the reference strong coupling $\tilde{\alpha}_s \equiv \alpha_s(M_Z)$. We note that the notation in Eq.~\ref{eq:pdfs-repr} is overloaded
as it also represents the parton-to-hadron fragmentation functions, usually denoted $(\Delta) D_i$.

Omitting the dependence on the atomic mass number $A$, the strong coupling $\tilde{\alpha}_s$, the polarisation, and assuming
\enquote{azimuthal symmetry}~\cite{Bastami:2018xqd,Anselmino:2013lza,Anselmino:2011ch}, the collinear PDF is recovered by integrating over the
parton's transverse momentum:
\begin{equation}
	f_i(x, Q^2) = \int 2 \pi k_T \, \mathrm{d} k_T f_i \left( x, k_T, Q^2  \right)
	\label{eq:unpol-pdfs}
\end{equation}
which has to satisfy the conservation of momentum and baryon number:
\begin{align}
	& \sum_{i} \int_{0}^{1} \mathrm{d}x \, x f_i(x, Q^2) = 1, \label{eq:pdf-cons-momentum} \\
	& \int_{0}^{1} \mathrm{d}x \left[ f_i(x, Q^2) - \bar{f}_i(x, Q^2) \right] = n_i,
	\label{eq:pdf-cons-baryon}
\end{align}
where $i$ runs over all the parton flavours and $n_i$ depends on the type of parton and hadron. For protons, $n_d = 1$, $n_u = 2$, and
$n_i = 0$ for $i \in \left\{  s, c, b, t \right\}$. It is evident from Eqs.~\ref{eq:pdf-cons-momentum} and \ref{eq:pdf-cons-baryon} that those constraints
hold for all values of the energy scale $Q^2$.

We must note that the validity of Eq.~\ref{eq:unpol-pdfs} is not always guaranteed. The definition of TMDs involves soft-factor
subtractions and rapidity regulators that render the operator matrix elements finite and well-defined in
QCD~\cite{Collins:2017oxh,Aybat:2011zv}. These subtractions, however, introduce scheme-
and scale-dependence that differ from the Minimal-Subtraction ($\overline{\rm MS}$) prescription used in collinear PDFs. As a consequence, the
naive $k_T$-integral of a TMD typically exhibits ultraviolet (UV) divergences and does not coincide with the $\overline{\rm MS}$ PDFs.

Different sum rules also exist for TMDs depending on the specific dynamics. For instance, for transversely polarised TMDs, the Sivers or Burkardt
Sum Rule~\cite{Burkardt:2005hp,Goeke:2006ef,Courtoy:2008dn} states that the net average transverse momentum of all partons due to the Sivers
effect vanishes when summed over all parton flavours.

Taking as a building block Eq.~\ref{eq:pdfs-repr}, the \neopdf~library provides a unified framework for constructing and interpolating PDFs and
related quantities. In contrast to traditional approaches where different classes of PDFs are treated within their own frameworks, \neopdf~is designed
to integrate in a consistent manner these varieties within a single coherent framework.
This new framework could pave the way to systematic studies of hadron structure across different kinematic regimes and external parameters, providing
a robust foundation for precision analyses at the (HL-)LHC, the upcoming EIC~\cite{AbdulKhalek:2021gbh,Abir:2023fpo}, and
beyond~\cite{FCC:2018byv,FCC:2018evy,FCC:2018vvp,InternationalMuonCollider:2025sys}.

\subsection{Features}
\label{subsec:features}

\neopdf~comes with a variety of features that ought to overcome the physics and technical limitations of existing interpolation libraries. In the following
part, we list some of the notable ones.

\paragraph{PDF classification.} Interpolated PDFs are mainly used to convolve with hard partonic cross-sections in order to calculate theoretical
predictions that could be compared to experimental measurements. The way in which convolutions are performed can be done in two ways: the
convolutions are calculated on the fly when the Monte Carlo generators are running; or the Monte Carlo weights are stored in some fast-interpolation
grids~\cite{Carrazza:2020gss,Carli:2010rw,Kluge:2006xs,Wobisch:2011ij,Britzger:2012bs} that could a posteriori be convolved with any PDFs and
different input settings. The latter provides several advantages, especially given the computational costs required to produces state-of-the-art
theory predictions.

Assume a process in which a single neutral Pion is produced from proton-proton collision, $p + p \to \pi^0 + X$. The corresponding
interpolation grid requires three distinct convolutions: two unpolarised PDFs for the protons and an unpolarised FF for the neutral Pion. When users
convolve such a grid with the three functions, they must pass them in the correct order to avoid calculating wrong predictions. This approach is, however, flawed
and error-prone.

\neopdf~addresses this issue by explicitly storing the complete attributes of a given distribution into the keys \codify{hadron\_pid}, \codify{polarised},
and \codify{set\_type}. The first key stores the PID value of the hadron, the second stores a boolean whose value is \codify{true} if the set is polarised,
and the last one specifies whether the PDF set is spacelike or timelike. These attributes can therefore be checked during the convolution to ensure that
the correct operations are performed.

\paragraph{No-code migration.} A decisive factor for the adoption of any new computational framework in high-energy physics is its compatibility with
existing workflows. Over the past two decades, \lhapdf~has become the de facto standard for accessing PDFs in both experimental analyses and theoretical
studies. Consequently, countless analysis pipelines, event generators, and fitting frameworks have been built around the  \lhapdf~interface. Any requirement
for extensive reimplementation would therefore pose a significant barrier to the deployment of a next-generation library, regardless of its technical superiority.

To address this challenge, \neopdf~has been designed as much as possible with a \enquote{no-code migration} philosophy across the various API interfaces.
This principle entails preserving naming conventions and method signatures in close alignment with \lhapdf, ensuring that existing codes can switch to 
\neopdf~with minimal or no modifications. In practice, functions for initialisation, member selection, and PDF evaluation are provided with identical calling
conventions to those of \lhapdf, allowing users to replace the underlying library in their build environment without altering analysis code, thereby safeguarding
interoperability with widely used software tools. Switching from \lhapdf~to \neopdf~thus should only require changing the package or module import. 

The benefits of such design compatibility are twofold. First, it removes barriers to adoption so that experimental collaborations and phenomenology groups can
immediately leverage \neopdf’s enhanced performance, extended interpolation capabilities, and reduced storage requirements without the need to refactor large
codebases. This is especially critical in high-stakes environments such as LHC data analyses, where stability and reproducibility are paramount. Second, it provides
backward compatibility with legacy PDF sets. By mimicking the \lhapdf~access patterns,  \neopdf~can act as a drop-in replacement, ensuring that legacy grids
can still be queried alongside new-generation \neopdf~sets within the same workflow.

\neopdf's no-code migration design removes the traditional friction associated with transitioning to new software infrastructures and ensures that its adoption
can be done in a seamless manner.

\begin{figure}[!tb]
	\centering
	\includegraphics[width=0.45\textwidth]{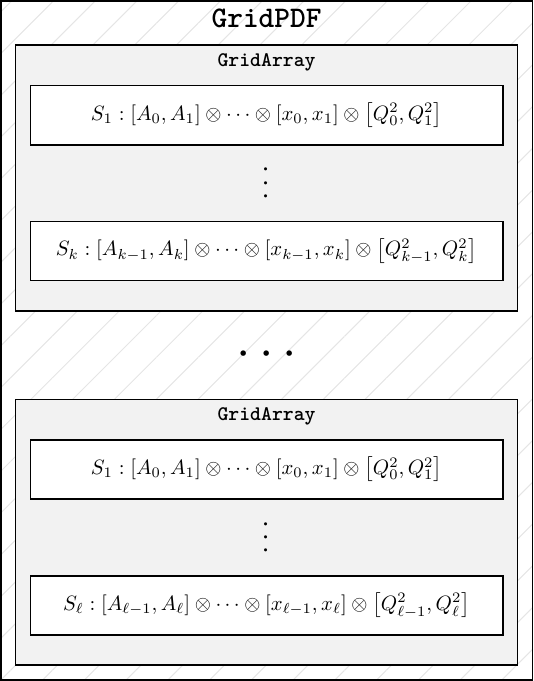}
	\caption{A diagrammatic representation of the \neopdf~data structure. The \texttt{GridPDF} object represents an instance of a given PDF set. It
	can contain multiple members, represented by the \texttt{GridArray} object. Each \texttt{GridArray} in turn contains a single or multiple
	subgrids---represented by the \texttt{SubGrid} object---which hold the information on the grid knots and values. Each subgrid $S_k$ is
	hyperrectangle defined by the Cartesian product of $N$ one-dimensional intervals, with $N$ representing the number of dependent variables.}
	\label{fig:gridarray}
\end{figure}

\paragraph{$A$-interpolation.} The need for interpolating along the atomic mass number $A$ is twofold. On the one hand, global nPDFs extraction
frameworks such as nNNPDF~\cite{AbdulKhalek:2019mzd,AbdulKhalek:2020yuc,AbdulKhalek:2022hcn} require the definition of so-called $t_0$ input
nPDF sets. For a given dataset, the input nPDF set is used to construct the definition of the $t_0$ covariance matrix entering the fitting
procedure to avoid the so-called \emph{D'Agostini Bias}~\cite{Agostini,NNPDF:2021njg,NNPDF:2021uiq}. Given that different
experimental datasets correspond to different nuclear targets, multiple nPDF grids must be loaded simultaneously. Therefore, the ability to consistently
interpolate across the atomic mass number $A$ within one single PDF grid instance removes the need to load multiple sets and streamlines the fitting
procedure.

On the other hand, extractions of nuclear parton densities may provide distribution functions for selected nuclei while experimental observables could
involve intermediate atomic mass numbers that are not explicitly included in the fitted sets. Conversely, there exist classes of nuclei, such as short-lived
isotopes ($^3$H) and exotic light nuclei ($^7$He, $^{11}$Li), which despite being of significant theoretical interests could not be accessed experimentally
 because they are either short-lived, unstable, or less abundant. In such cases, having a unified framework to represent the varieties of nuclear PDFs
 and being able to interpolate on them represents a critical step toward achieving consistent and high-precision QCD phenomenology in the nuclear
 sector.
 
 \paragraph{$\tilde{\alpha}_s$--interpolation.} Of equal relevance is \neopdf's ability to interpolate along the strong coupling $\tilde{\alpha}_s = \alpha_s(M_Z)$.
 Traditionally, global fitting groups only release small number of PDF sets corresponding to specific values of $\tilde{\alpha}_s$. This limitation could
 complicate the propagation of theoretical uncertainties associated with $\tilde{\alpha}_s$, especially in precision collider predictions where variations of
 $\Delta \tilde{\alpha}_s \sim  10^{-4}$ are needed.
 By incorporating $\tilde{\alpha}_s$ as an explicit interpolation dimension, \neopdf~provides continuous access to variations of the strong coupling while preserving
 its correlations with the PDFs.
 
 \paragraph{Modular interpolation strategies.} In legacy frameworks, the interpolation is tightly coupled to the data model and provided as a
 fixed scheme. While this may guarantee stability, it can limit the ability to exploit recent advances in numerical methods or to tailor the interpolation strategy
 to specific phenomenological requirements.
 
 \neopdf~addresses this shortcoming by explicitly disentangling the interpolation routine from the data structure. Within its design, the grid storage and access
 layers are agnostic to the choice of interpolation, therefore enabling users to seamlessly introduce custom strategies without altering the core library.
 
 This modularity yields several practical advantages. For instance, the decoupling of the interpolations from the underlying data structure ensures that numerical
 methods can be adapted without compromising on compatibility or reproducibility. Furthermore, it allows PDF fitting groups to optimise their sets according to
 some custom interpolation schemes.

\subsection{Grid representation}
\label{subsc:grid-representation}

In order the tackle the growing complexity of hadron description, \neopdf~introduces a modern subgrid-based design that avoids the rigid limitations
of existing grid systems, while ensuring flexibility, efficiency, and scalability.

The core concept of the \neopdf's data structure is diagrammatically illustrated in Fig~\ref{fig:gridarray}. At the top level, a given PDF set is represented
by a \texttt{GridPDF} object, which serves as the container for all information related to that set. A \texttt{GridPDF} may include multiple members, each
stored as an independent \texttt{GridArray}. The \texttt{GridArray} in turn encapsulates the full kinematic information by subdividing the interpolation space
into elementary building blocks called subgrids. These subgrids form the lowest-level representation of the distribution, holding the explicit grid knots and
corresponding values.

The subgrid structure provides the fundamental building block for storing and accessing parton distributions across multiple kinematic variables. Each
subgrid hence represents a restricted domain in the space of interpolation variables. In the most general case, for a given parton flavour, a subgrid is represented
as a rank-5 tensor:
\begin{equation}
	\mathfrak{S}_{i, j, k, \ell, m, n} \equiv \mathfrak{S} \left( A_i, \tilde{\alpha}_{s, j}, k_{T, \ell}, x_m, Q^2_n \right),
	\label{eq:gridarray}
\end{equation}
whose dimensions correspond to the atomic mass number $A$, the strong coupling $\tilde{\alpha}_s$, the intrinsic transverse momentum $k_T$, the
longitudinal momentum fraction $x$, and the hard scale $Q^2$. Depending on the distribution, this general structure naturally reduces to lower-dimensional
arrays. For instance, in the standard case of proton PDF, it reduces to a 2-dimensional array of $\mathfrak{S}_{m, n}$. 

The flexibility of the subgrid formalism lies in its subdivision of the full interpolation range into smaller intervals, each covering a restricted domain
$\left[ z_{i-1}, z_i \right]$ for a given variable $z \in \left\{ A, \tilde{\alpha}_s, k_T, x, Q^2 \right\}$. The full \texttt{GridArray} hence is a conjunction of $k$
subgrids that can be represented as:
\begin{equation}
	\left[ z_0, z_1, \cdots, z_k  \right]_{\left( n_1, n_2, \cdots, n_k \right)},
\end{equation}
where each interval representing a subgrid $S_i$ is discretized with $n_i$ interpolation points. Note that given adjacent subgrids
share their end-points, the total number of grid points for a given variable is:
\begin{equation}
	n_{\rm pts } = \sum_{i = 0}^k n_i - (k - i).
\end{equation}
It thus follows that the total number of subgrid for a given \texttt{GridArray} is the Cartesian product of the subdivisions chosen for the each interpolation
variable. Taking the 2-dimensional subgrid example of the proton PDF, if the interpolation domain is partitioned into $k_x$ intervals in $x$ and $k_{Q^2}$
intervals in $Q^2$, then the total number of subgrids is $N_{\rm subgrids} = k_x \times k_{Q^2}$.

This systematic subdivision provides the essential scalability of \neopdf, enabling it to accommodate standard collinear proton PDFs to fully
differential multidimensional distributions within a unified and consistent framework.

\subsection{Interpolation strategies}
\label{subsec:interpolation}

\begin{figure*}[tb]
	\centering
	\includegraphics[width=0.495\textwidth]{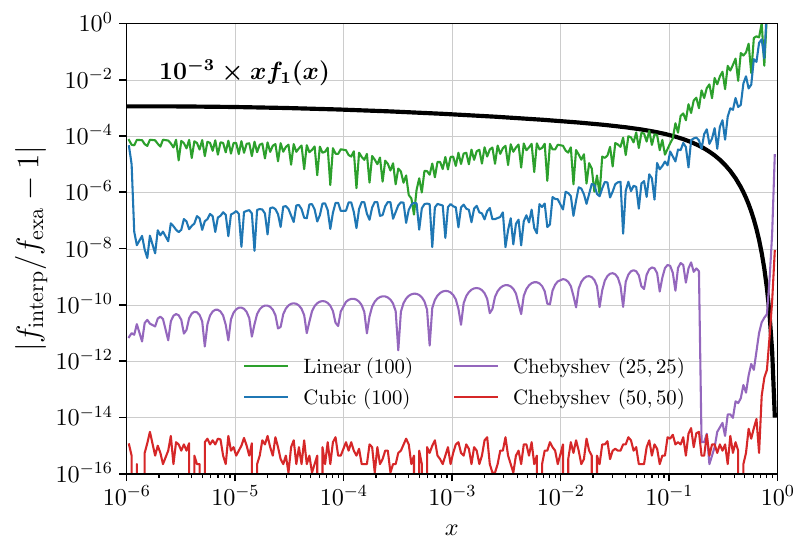}
	\includegraphics[width=0.495\textwidth]{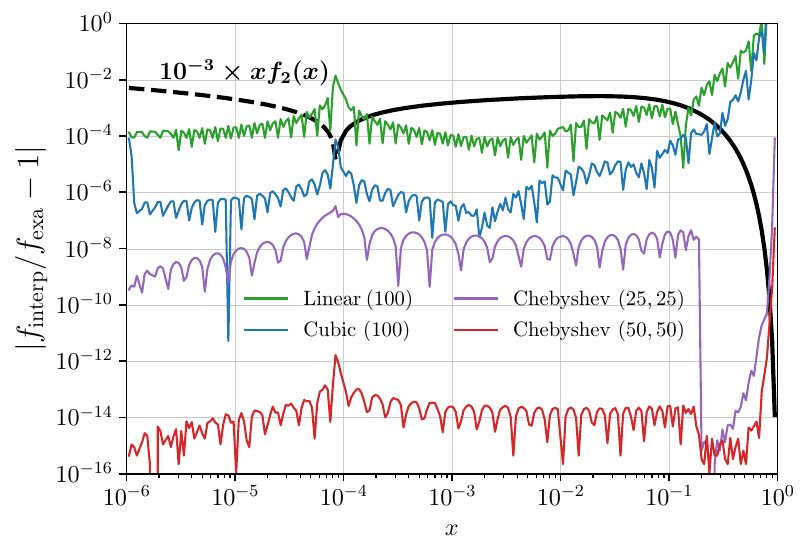}
	\caption{Relative interpolation accuracy for the two representative PDFs give in Eqs.~\ref{eq:abmp} and \ref{eq:herepdf20}.
	The linear (green) and cubic spline (blue) interpolants both use the same total number of grid points $n_{\rm pts} = 100$. The Chebyshev interpolants
	are given with both low- and high-density grids, $n_{\rm pts} = 49$ (purple) and $n_{\rm pts} = 99$ (red) respectively. While the linear and cubic spline interpolants
	are expressed in terms of a single subgrid $\left[ 10^{-6}, 1 \right]_{(100)}$, the Chebyshev counterparts are represented in terms of two subgrids
	$\left[ 10^{-6}, 0.2, 1 \right]_{(n, n)}$ where $n=25,50$ depending on the density of the grid. The exact results that are being interpolated are shown
	in (dashed) black.}
	\label{fig:chebyshev}
\end{figure*}

A key design principle of  \neopdf~is that the choice of interpolation method is a configurable property defined directly within the
metadata of the grid itself. This is to make sure that the grid knots spacing is consistent with the chosen interpolation. In \neopdf,
each grid contains a metadata with a field called \codify{InterpolatorType}. When the library loads a grid, it parses this field and
uses its value to instruct the \codify{InterpolatorFactory} on which strategy to instantiate. This approach renders the grid data
self-describing and provides significant flexibility while allowing PDF fitting groups to choose the most appropriate interpolation
algorithm for their PDF determination.

The interpolation in \neopdf~is constructed through a dynamic, multi-strategy approach. When a PDF is queried, the \codify{InterpolatorFactory}
inspects each subgrid to determine its dimensionality, whether it is two-dimensional in $x$ and $Q^2$, or three-/four-/five-dimensional if
$A$, $\tilde{\alpha}_s$, or $k_T$ variations are present. Based on the dimensionality and the \codify{InterpolatorType} specified in the metadata,
the factory creates and configures the corresponding interpolator for each flavour in that subgrid. Finally, when a value is requested, the call
is dispatched to the pre-built interpolator for the corresponding kinematic region.

In the current version, \neopdf~implements the following interpolation strategies:
\begin{itemize}
	\item[-] \textbf{Linear}: a linear interpolation for two-dimensional data with (\codify{LogBilinear}) and without (\codify{Bilinear}) logarithmic scaling
	of the coordinates. A $n$-dimensional version (\codify{InterpNDLinear}) is also provided to interpolate four- and five-dimensional data, which arises
	when the nucleon number $A$ and/or the strong coupling $\tilde{\alpha}_s$ are involved.
	\item[-] \textbf{Cubic Hermite Splines}: a cubic interpolation strategy for two-dimensional (\codify{LogBicubic}) and three-dimensional (\codify{LogTricubic})
	data with logarithmic scaling of the input coordinates. The two-dimensional implementation is exactly the same as the one implemented
	within LHAPDF while the three-dimensional implementation is an extension of it.
	\item[-] \textbf{Chebyshev}: a global interpolation strategy using Chebyshev polynomials. Its implementation (\codify{LogChebyshev})
	supports one-, two-, and three-dimensional data.
\end{itemize}

The use of Chebyshev polynomials to interpolate PDFs has been proposed in~\cite{Diehl:2021gvs} to significantly improve numerical accuracies when
compared to cubic splines approaches at lower computational costs. The idea is to approximate a given function over its entire domain
using a non-equidistant knots that clusters the grid points toward the edges of the interval. For piecewise interpolations such as cubic
splits, the error polynomially decreases as a function of the number of grid points, while for Chebyshev interpolations, it decreases
exponentially.
Furthermore, Chebyshev interpolants are $C^{\infty}$ continuous, and therefore yields numerically stable derivatives. Such a property
is crucial when studying subleading power corrections, where the first and second derivatives of PDFs with respect to $x$ are explicitly
required; or when analysing the asymptotic behaviour of the PDFs at small- and large-$x$ which necessitates the explicit calculations of
the logarithmic derivates of the PDFs.

To illustrate the efficiency of the Chebyshev interpolation for PDFs, we compare its accuracy with linear and cubic Hermite splines.
As implemented in \neopdf, these interpolations perform logarithmic scaling of the input coordinates. We benchmark the accuracies of
the interpolations by considering two representative test functions:
\begin{align}
	xf_1(x) &= c_1 \exp\left[\left( 1 + \beta_1 x \right) \left( \ln x + \beta_2 \ln^2 x \right) \right] \left( 1 - x  \right)^{\beta_3} \label{eq:abmp} \\
	xf_2(x) &= c_1 x^{\beta_1} \left( 1 - x  \right)^{\beta_2} - c_2 x^{\beta_3} \left( 1 - x \right)^{\beta_4}, \label{eq:herepdf20}
\end{align}
which with the right coefficients correspond to PDFs at the input scale of different common PDF sets. Specifically, $f_1$ represents the
parametrisation of the anti-up quark PDF $f_{\bar{u}}$ from the ABMP16~\cite{Alekhin:2018pai} extractions at next-to-next-to-leading (NNLO)
order in perturbative accuracy, while $f_2$ represents the parametrisation of the gluon PDF $f_g$ from the
HERAPDF20~\cite{H1:2015ubc} extraction at next-to-leading order (NLO).

We note that for the Chebyshev interpolation strategy to work, the function to be interpolated has to be computed specifically on the Chebyshev
nodes. That is, for a given $z_j$ values defined within the domain $\left[ z_{\rm min}, z_{\rm max} \right]$ with $n$ points, they are mapped onto:
\begin{equation}
	\tilde{z}_j=\exp \left[u_{\rm min}+\frac{u_{\rm max}-u_{\rm min}}{2}\left(1+\cos \left(\frac{\pi(n-1-j)}{n-1}\right)\right)\right]
\end{equation}
where $j = 0, \cdots, n-1$, and  $u_{\rm min} = \ln \left( z_{\rm min} \right)$ and $u_{\rm max} = \ln \left( z_{\rm max} \right)$ are the affine maps
to the log-space interval.

Fig.~\ref{fig:chebyshev} shows the relative interpolation accuracy for the two representative functions $xf_1$ (left) in Eq.~\ref{eq:abmp} and
$xf_2$ (right) in Eq.~\ref{eq:herepdf20}. The exact functions that are being interpolated are shown by the black thick solid and dashed lines for
positive and negative values, respectively. The linear (green) and cubic spline (blue) interpolants are both expressed in terms of a single
subgrid $\left[ 10^{-6}, 1 \right]_{(100)}$ and use the same total number of grid points $n_{\rm pts} = 100$. The Chebyshev interpolant is instead
expressed in terms of two subgrids $\left[ 10^{-6}, 0.2, 1 \right]_{(n, n)}$ where $n$ varies depending on the density of the grid. We consider
a high-density grid with $(50, 50)$-point and a low-density version with $(25, 25)$-point, each containing a total number of grid points
$n_{\rm pts} = 99$ and $n_{\rm pts} = 49$, respectively.
As expected, the cubic spline is more accurate than the linear interpolation, especially at low- and medium-$x$ ($x \lesssim 10^{-2}$), with the
cubic spline reaching a relative accuracy of $10^{-6}$. However, both of these interpolation strategies significantly deteriorates as $x \sim 1$
with a relative interpolation accuracy worse than $10^{-2}$. With the same number of total grid knots, the Chebyshev interpolation outperform
both interpolations by several orders of magnitudes. What is perhaps more impressive is the interpolation accuracy at large-$x$, reaching beyond
$10^{-2}$ until it approaches the non-analytic region $x=1$. This could mainly be explained by the fact that the second subgrid, which contains half
of the grid points, is within the domain $\left[ 0.2, 1 \right]$.
We also see that even with half the number of grid points, the Chebyshev interpolation still outperforms both the linear and cubic spline interpolations
by a few orders of magnitude. This demonstrates that interpolation accuracy scales far more efficiently with the number of points in the Chebyshev
interpolation than with piecewise-based methods.

Because the Chebyshev interpolation strategy requires substantially fewer number of grid points to achieve excellent accuracies, it yields
more compact grids with smaller file sizes, which is particularly relevant for large and dense ensembles. Therefore, while cubic splines remain
suitable for legacy \lhapdf~sets, Chebyshev interpolations are the recommended choice for \neopdf~grids.

\subsection{Extrapolation strategies}
\label{subsec:extrapolation}

The default extrapolation in \neopdf~is to \enquote{freeze} the PDF interpolations at the boundaries. In principle, some sensible extrapolation
strategies could be implemented; some examples include the MSTW \enquote{continuation} extrapolation~\cite{Buckley:2014ana,Martin:2009iq}, which
is the default extrapolator in \lhapdf. However, we leave the exploration of such strategies for future works.

We would like to emphasize that because of the subgrid-system in \neopdf, finding the closest subgrid for a given input coordinate is now more
involved. In the case a point does not lie inside any subgrid, the algorithm to select the appropriate subgrid relies on computing the Euclidean distance
of the point to all the subgrids.

Specifically, let $P$ denote a point in an $N$-dimensional parameter space and $\left\{ S_j \right\}^{k}_{j=1}$ be the set of subgrids. As introduced in
Sec.~\ref{sec:design}, each subgrid $S_j$ is defined as a hyperrectangle given by
the Cartesian product of the intervals:
\begin{equation}
	S_j = \left[ A_{j-1}, A_{j} \right] \otimes \cdots \otimes  \left[ x_{j-1}, x_j \right] \otimes  \left[ Q^2_{j-1}, Q^2_j \right] 
\end{equation}
The distance from the point $P$ to a subgrid $S_j$ is thus the Euclidean distance from $P$ to the closest point within $S_j$. For efficiency, the
implementation uses the squared Euclidean distance, as it preserves the ordering of distances without requiring costly square root calculations.
The squared distance between $P$ and a subgrid $S_j$ is computed component-wise along the variables. For a given variable $z$, the contribution
is given by:
\begin{equation}
	d\left(p_z,\left[z_{j-1}, z_j\right]\right)^2= \begin{cases}
		\left(z_{j-1}-p_z\right)^2 & \mathrm{if} \quad p_z<z_{j-1}, \\
		\left(p_z-z_j\right)^2 & \mathrm{if} \quad p_z>z_j .
	\end{cases}
\end{equation}
The total squared Euclidean distance for the $j$-th subgrid is thus the sum of the distances over all the dimensions:
\begin{equation}
	\mathcal{D}^2 \left( P, S_j \right) = \sum_{z = A, \cdots, x, Q^2 } d\left(p_z,\left[z_{j-1}, z_j\right]\right)^2 .
	\label{eq:distance}
\end{equation}
By iterating through the subgrids and computing Eq.~\ref{eq:distance}, the point closest to the subgrid in the Euclidean metric is defined as follows:
\begin{equation}
	s^\star = \arg \min _{j \in\{1, \ldots, k\}} \mathcal{D}^2\left(P, S_j\right).
\end{equation}
This procedure guarantees that extrapolation is performed relative to the physically most relevant region of the parameter space, ensuring continuity
across all adjacent subgrid boundaries.

\subsection{\neopdf~data format}
\label{subsec:format}

A core component of the \neopdf~framework lies in its dedicated file format, specifically engineered to overcome the rigidity and inefficiencies of legacy
interpolation libraries. As modern QCD phenomenology demands the description of increasingly complex distributions, the storage format must provide
not only numerical precision and compactness, but also flexibility, scalability, and efficient access patterns.

The \neopdf~format achieves these goals by combining binary serialisation with LZ4 compression, thereby reducing significantly storage requirements. In
addition to compactness, this design ensures rapid I/O performance, which is essential when handling large ensembles of PDF sets. The choice of a binary, 
non-human-readable format reflects a deliberate emphasis on efficiency, where metadata and offset information provide the necessary transparency for
programmatic access. To distinguish from standard PDF sets, \neopdf~grids are suffixed with \texttt{.neopdf.lz4}. To compensate for the non-human readable
format, \neopdf~provides a Command Line Interface (CLI) tool, allowing users to perform PDF interpolation, $\alpha_s$ evaluation, file format conversion, and
metadata inspection directly from the terminal.

At the structural level, a \neopdf~grid contains a metadata header, which is immediately accessible without decompression. This header contains all essential
information on the set, including the hadronic target, the class of distributions, the kinematic variables, and other metadata. Following the header, an offset table
provides direct pointers to the data blocks of individual members, enabling random access without the need to sequentially traverse the file. The main body of
the file is composed of grid data blocks, each corresponding to a \texttt{GridArray}. These blocks are individually serialised, compressed, and length-prefixed to
facilitate fast streaming and accurate deserialisation. Formally, this may be represented as a mapping:
\begin{equation}
	\mathcal{F}=\left\{\mathcal{H}, \mathcal{O}, \bigcup_{m=0}^{N_{\text {mem }}} \mathcal{G}_m\right\},
\end{equation}
where $\mathcal{H}$ denotes the metadata header, $\mathcal{O}$ the offset table, and $\mathcal{G}_m$ the compressed grid data block for the $m$-th
member. As detailed in previous sections, each gird block itself is a collection of subgrids. The offset table $\mathcal{O}$ thus defines a bijection:
\begin{equation}
	\mathcal{O}: m \longmapsto \mathrm{addr} \left(\mathcal{G}_m\right),
\end{equation}
that maps a member $m$ to the address of the corresponding grid, allowing for a direct random access to each block.

This structure allows for several efficient modes of access. First, the entire sets may be \underline{eagerly} loaded into memory when analyses
require simultaneous evaluation of many members. Alternatively, specific grids can be retrieved on demand via the offset table, allowing fast
random access even in very large collections. Finally, for memory-constrained environments, \underline{lazy} iteration provides a further option,
enabling sequential traversal of members with minimal resource usage.

\begin{table}[!h]
	\renewcommand{\arraystretch}{1.60}
	\setlength{\heavyrulewidth}{0.08em}
	\setlength{\lightrulewidth}{0.05em}
	\setlength{\cmidrulewidth}{0.03em}
	\scriptsize
	\centering
	\begin{tabularx}{0.475\textwidth}{llXX}
		\toprule
		PDF Set & Nb. Members & \lhapdf/\tmdlib & $\quad$\neopdf \\
		\midrule
		PDF4LHC21 & 40 & 31 MB & $\quad$16 MB \\
		\midrule
		NNPDF4.0 NNLO & 100 & 158 MB & $\quad$85 MB \\
		\midrule
		NNPDF4.0 NNLO & 1000 & 1.55 GB & $\quad$830 MB \\
		\midrule
		Combined nNNPDF3.0 & 19$\times$200 & 19$\times$160 MB & $\quad$1.43 GB \\
		\midrule
		MAP22 FF @N3LL & 250 & 2.50 GB & $\quad$1.25 GB \\
		\bottomrule
	\end{tabularx}
	\vspace{0.3cm}
	\caption{File size comparison between the legacy \lhapdf~and~\tmdlib~formats and the \neopdf~format for representative PDF sets. The table demonstrates that
	\neopdf~typically reduces storage requirements by approximately half.}
	\label{tab:set-sizes}
\end{table}

The advantages of the \neopdf~file format are threefold. First, it provides a drastic gain in performance, as binary serialisation combined with LZ4 compression
ensures both compact storage (see Tab.~\ref{tab:set-sizes}) and fast read–write cycles. Second, it guarantees scalability, since the access patterns are optimized
to handle increasingly large and diverse PDF sets, including those that explicitly stores the $A$, $\tilde{\alpha}_s$, and $k_T$ dependence. Finally, the design is inherently
extensible: new metadata fields, additional kinematic dimensions, or even novel classes of hadronic distributions can be accommodated without breaking compatibility.

\section{Benchmark against other interpolation libraries}
\label{sec:benchmark}

In order to assess the performance and reliability of the \neopdf~framework, we conduct a systematic benchmark against the \lhapdf~and \tmdlib~libraries
for various PDF settings. The comparisons focus on three key aspects: the efficiency of loading PDF sets, the numerical accuracy of the
interpolated results, and the computational speed of the  interpolations across relevant kinematic ranges.

\subsection{\lhapdf}
\label{subsec:benchmark-lhapdf}

We start by benchmarking \neopdf~against \lhapdf~for collinear parton distributions using the same cubic spline interpolation. We demonstrate
that \textsc{NeoPDF} not only ensures perfect compatibility with \lhapdf, but also offers substantial improvements. We note that, unless otherwise
specified, we always perform the benchmark using the Python APIs of both libraries.

\begin{figure}[!h]
	\centering
	\includegraphics[width=0.495\textwidth]{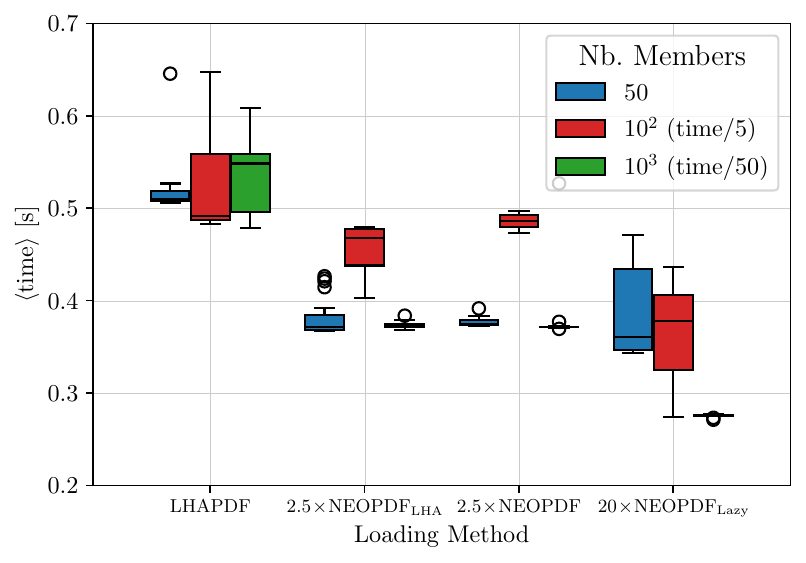}
	\caption{Loading time for PDF sets in \lhapdf~and \neopdf, averaged over 20 independent runs, as a function of the number of members
	in the ensemble. Results are shown for sets of $50$, $10^2$, and $10^3$ members. For visualisation purposes, the times for the latter
	two sets are rescaled by factors of $1/5$ and $1/50$, respectively. Similarly, the eager (\neopdf$_{\rm LHA}$ and \neopdf) and lazy
	(\neopdf$_{\rm Lazy}$) loading methods have been scaled by factors of $2.5$ and $20$, respectively.}
	\label{fig:bench-loading}
\end{figure}

\begin{figure*}[!tb]
	\centering
	\includegraphics[width=0.495\textwidth]{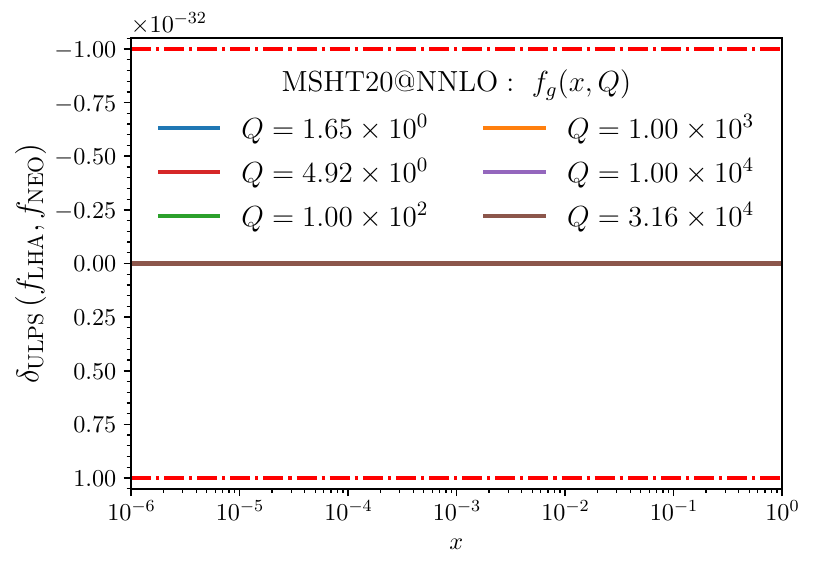}
	\includegraphics[width=0.495\textwidth]{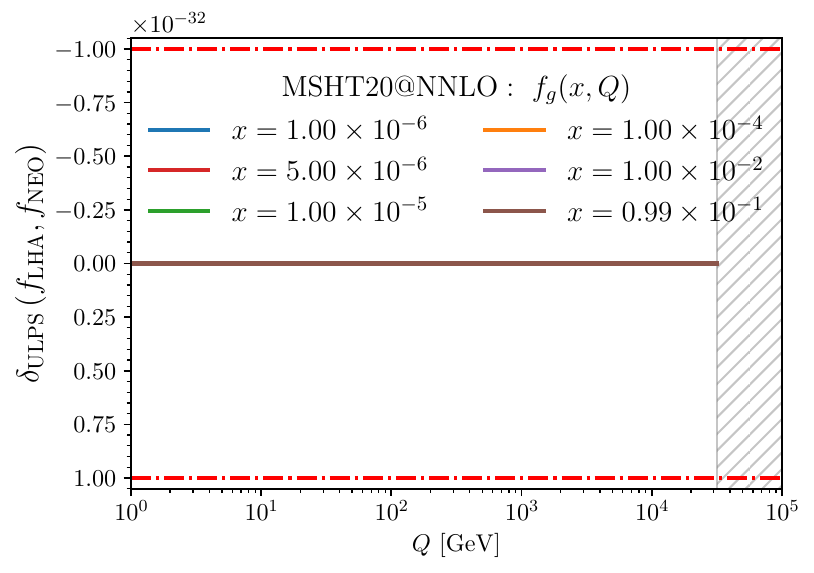}
	\caption{Interpolation accuracy of \neopdf~compared to \lhapdf, calculated using the Difference in Units of Last Places $\delta_{\rm ULPS}$.
	The results are shown for the MSHT20 NNLO set for the gluon distribution $f_g$ across the full $(x, Q)$--plane. The $x$ values of the MSHT20
	set spans the region $\left[10^{-6}, 1\right]$ while the hard scale $Q$ is defined within the domain
	$\left[1, 3.162 \times 10^{4}\right]~\mathrm{GeV}$.}
	\label{fig:lhapdf-bench-accuracy}
\end{figure*}

\paragraph{Data access.} The benchmark results for the loading time are shown in Fig.~\ref{fig:bench-loading}, where the average time over 20 independent
runs was measured as a function of the number of members in the set. The box shows the interquartile range (IQR) with a line representing the
median; the whiskers extend to the most extreme values within $1.5 \times$IQR of the quartiles, and the circles mark the outliers beyond that range.
The comparison highlights the different access strategies available in \neopdf, namely an eager loading of the entire set and a lazy
iteration over the members (\neopdf$_{\rm Lazy}$). The eager loading is shown both for the native \neopdf~input format (\neopdf) and for the
compatibility mode with \lhapdf (\neopdf$_{\rm LHA}$).
The results are shown for ensembles of $50$, $10^2$, and $10^3$ where, for visualisation purposes, the latter two are scaled by factors of $1/5$ and $1/50$, respectively.
We see that the eager loading in \neopdf~is significantly faster than the equivalent \lhapdf, with a typical reduction in loading time by a factor of at least 2.5 across the tested
ensemble sizes. The lazy loading mode further amplifies this advantage, providing up to an order of magnitude improvement.

It is worth emphasizing that the impressive reduction in the initialisation time observed for the lazy loading mode does not come without cost. Since members are only 
deserialised and decompressed when they are actually accessed, the price of lazy initialisation is effectively deferred to the interpolation stage. In other words, while the 
loading benchmark shows a significant advantage for \neopdf$_{\rm Lazy}$, the overall performance gain depends on the number of members that are subsequently queried. 
For scenarios that require only a small subset of replicas, the lazy strategy offers a substantial net improvement. Conversely, if all members are eventually accessed during
the interpolation, then the cumulative cost approaches that of eager loading, and the main advantage of the lazy mode is confined to memory management rather than
absolute speed.

\paragraph{Interpolation accuracy.} A critical validation of the \neopdf~library is the quantification of its interpolation accuracy relative
to the widely used \lhapdf. Given that one of the main aims of \neopdf~is to be an \lhapdf~alternative, it is crucial to make sure that---upon
using the exact same cubic spline interpolation---the results agree with remarkable precision.
To this end, we employ the Difference in Units of Last Places ($\delta_{\rm ULPS}$), which is a measure commonly used in numerical analysis to
quantify the discrepancies between floating-point numbers beyond conventional relative or absolute tolerances.

\begin{figure}[!h]
	\centering
	\includegraphics[width=0.495\textwidth]{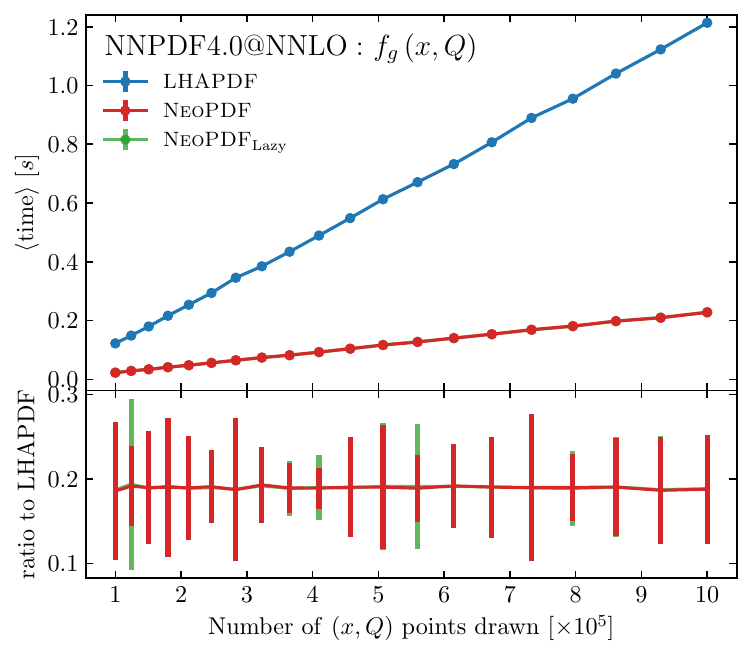}
	\caption{Benchmark of the interpolation evaluation time for the gluon distribution $f_g$ using the NNPDF4.0@NNLO set. The average wall-clock
	time per evaluation are shown as a function of the number of $(x, Q)$ points (top) together with the ratio relative to \lhapdf~(bottom).}
	\label{fig:lhapdf-bench-performance}
\end{figure}

\begin{figure*}[!tb]
	\centering
	\includegraphics[width=0.495\textwidth]{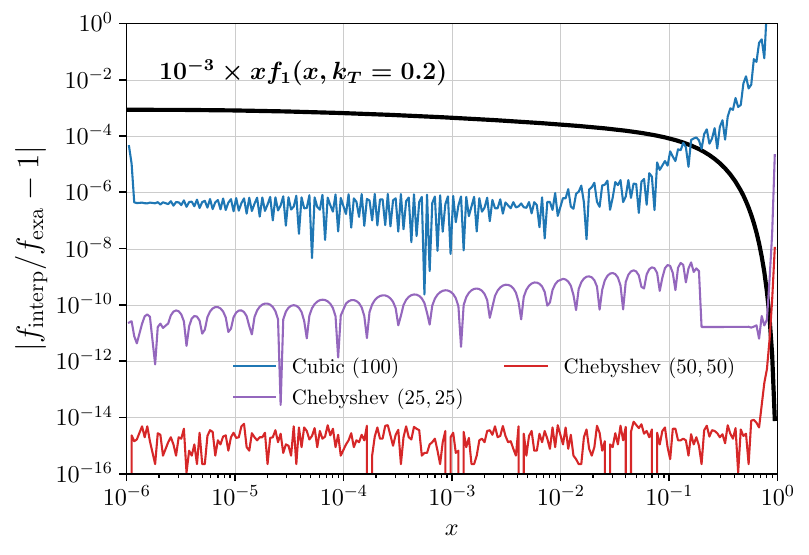}
	\includegraphics[width=0.495\textwidth]{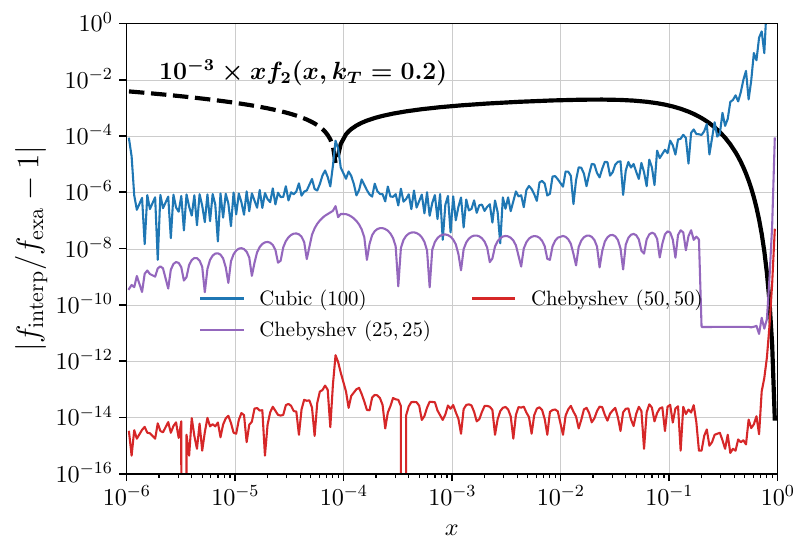}
	\includegraphics[width=0.495\textwidth]{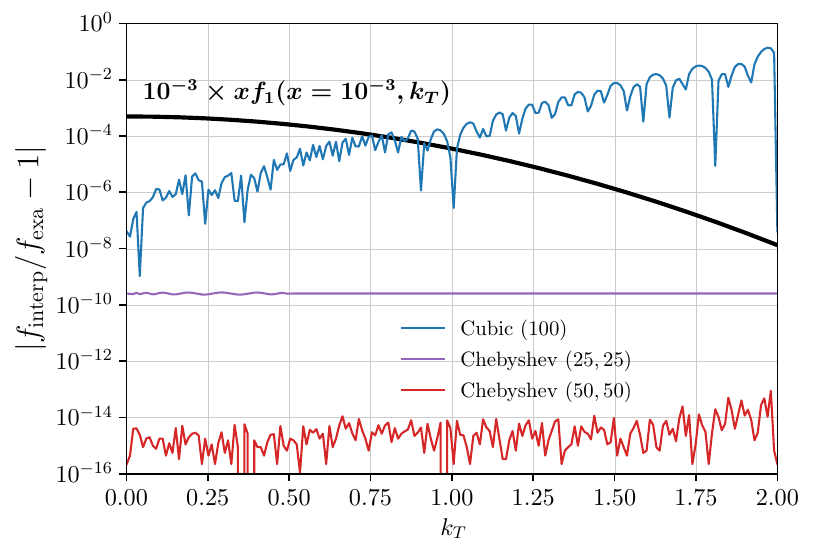}
	\includegraphics[width=0.495\textwidth]{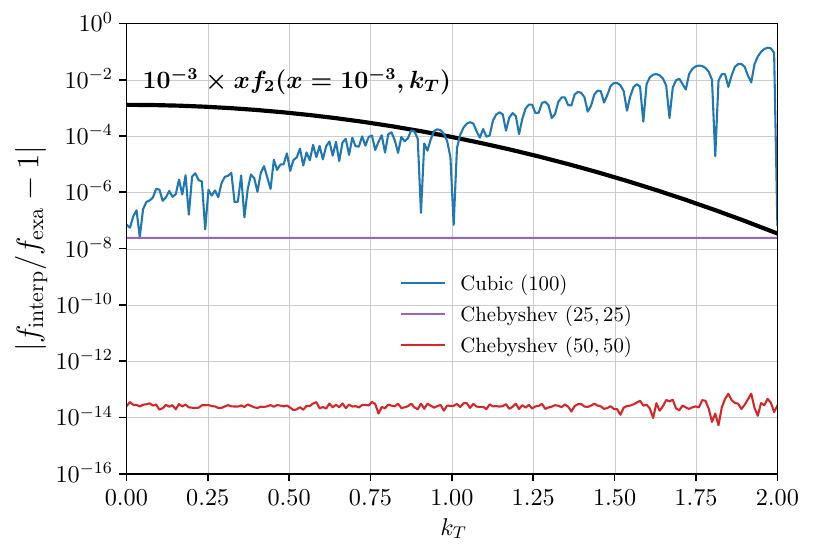}
	\caption{Similar to Fig.~\ref{fig:chebyshev} but for the representative TMD functions given in Eq.~\ref{eq:guassian-ansatz}. Shown are the results
	from the cubic (blue) and Chebyshev interpolations as a function of $x$ (top) and $k_T$ (bottom). The cubic spline interpolants always use a
	single subgrid with a total number of points
	$n_{\rm pts} = 100$ for both the longitudinal momentum fraction $x$ and the transverse momentum $k_T$. The Chebyshev interpolants
	use two subgrids for both $x$ and $k_T$, that is $\left[10^{-6}, 0.2, 1\right]_{(n, n)}$ and $\left[ 10^{-3}, 0.5, 2 \right]_{(n, n)}~\mathrm{GeV}$, respectively.
	The Chebyshev results are given for both a low- (purple) and high-density (red) grids. The exact results that are being interpolated are shown
	by the black solid and dashed lines for positive and negative values, respectively.}
	\label{fig:tmdlib-interp-accuracy}
\end{figure*}

Given two floating-point numbers $f_a, f_b \in \mathbb{F}_{64}$, with $\mathbb{F}_{64}$ denoting the set of IEEE-754 double-precision
floating-point numbers~\cite{4610935}, $\delta_{\rm ULPS}$ is defined as:
\begin{equation}
	\delta_{\rm ULPS} \left( f_a, f_b \right) = \lvert \mathrm{ULP}\left( f_a \right) - \mathrm{ULP}\left( f_b \right) \lvert,
\end{equation}
where $\mathrm{ULP}: \mathbb{F}_{64} \longmapsto \mathbb{Z}$ maps the set of floatting-point numbers to the set of signed integers that
represent their IEEE-754 encodings. In practice, $\delta_{\rm ULPS}$ counts the number of representable floating-point numbers between
two results, thus providing a direct measure of their bit-level agreement. For reference, a discrepancy of one-ULP corresponds to a relative
error of order $10^{-16}$, which is entirely negligible for phenomenological applications, while a value of $\delta_{\rm ULPS} = 0$ indicates
an exact equality in binary representation.

The benchmark results are shown in Fig.~\ref{fig:lhapdf-bench-accuracy}, where the MSHT20@NNLO set is tested across
the entire range in $x$ (left) and $Q$ (right) for the gluon PDF. Due to the differences in the implementation  of the extrapolation in both
libraries, the results are restricted to the set domains. We see from these results that $\delta_{\rm ULPS} = 0$ for the entire kinematic ranges,
demonstrating that the interpolated values obtained with \neopdf~agree with those of \lhapdf~beyond machine precisions.

These benchmark therefore show that when using the same cubic spline interpolation method, both libraries yield effectively identical results. 

\begin{figure*}[!tb]
	\centering
	\includegraphics[width=0.495\textwidth]{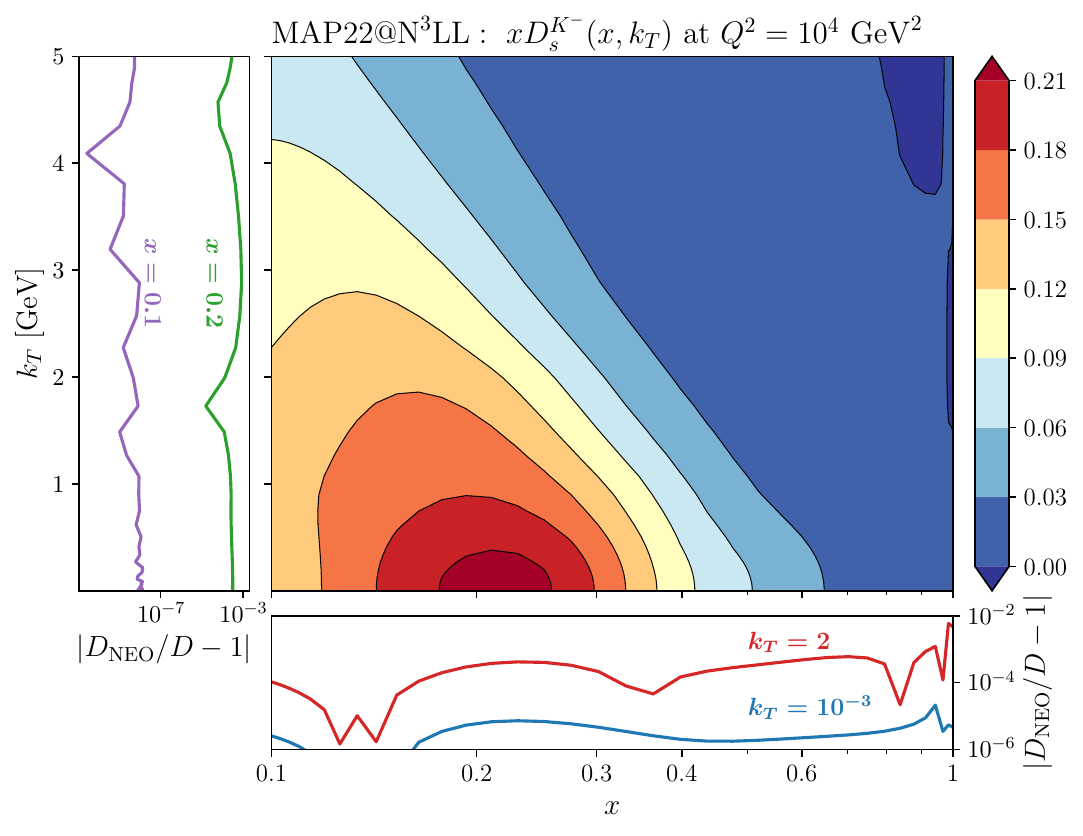}
	\includegraphics[width=0.495\textwidth]{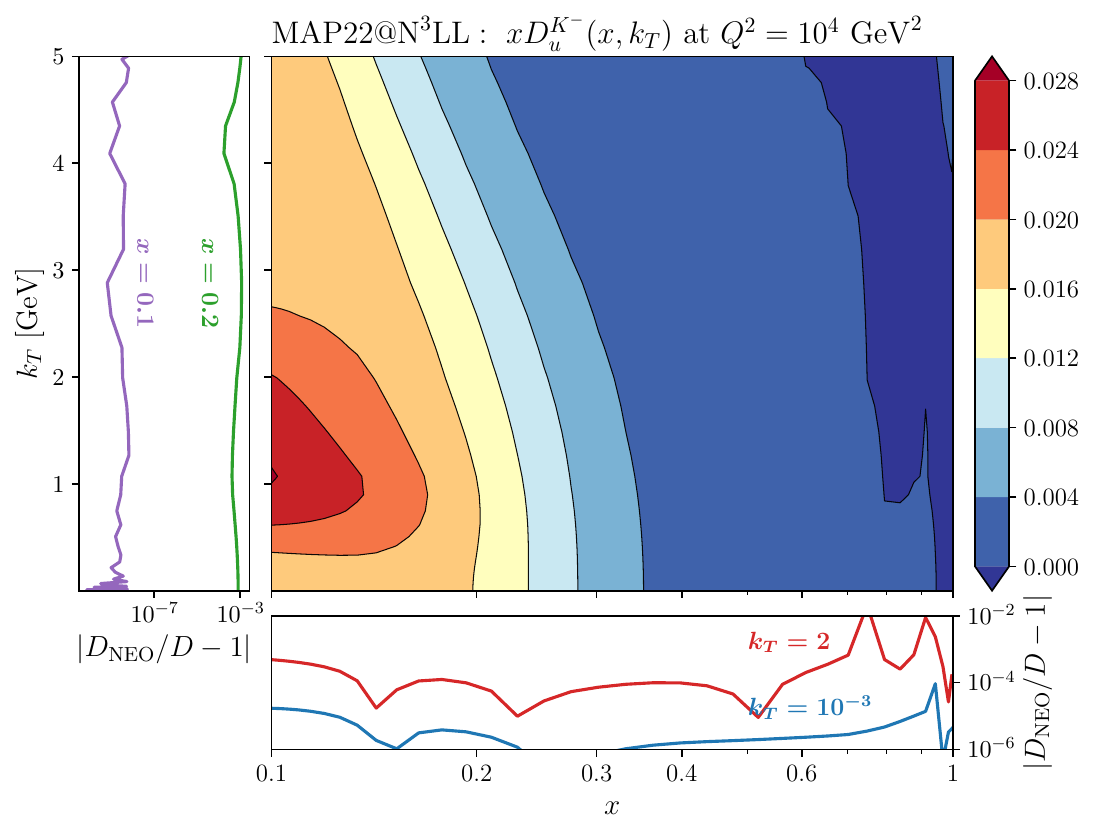}
	\caption{Benchmark of the transverse momentum-dependent parton-to-fragmentation functions comparing \neopdf~and \tmdlib. The benchmark were
	performed using the MAP22 FF set which includes next-to-next-to-next-to-leading logarithmic (N$^3$LL) corrections. The central panel shows the
	absolute values of the negatively charged Kaon TMD $D^{K-}_i(x, k_T)$ for the strange (left) and up (right) quark distributions across the entire range of
	transverse momentum $k_T$ and momentum fraction $x$, computed using \neopdf. The relative differences with respect to \tmdlib~are shown for fixed
	values of $x$ (left) and $k_T$ (bottom). Results are computed at a fixed value of the hard scale $Q^2 = 10^4~\mathrm{GeV}^2$.}
	\label{fig:tmdlib-bench-accuracy}
\end{figure*}

\paragraph{Interpolation evaluation time.}

Beyond accuracy, the efficiency of the interpolation is a decisive factor in the practical usability of any PDF interpolation framework. To
assess this aspect, we benchmark the evaluation time of \neopdf~agaisnt \lhapdf~using the NNPDF4.0@NNLO set. Specifically, we look at the gluon
distribution $f_g(x,Q)$. The benchmark measures the average wall-clock time required to evaluate the PDF for different numbers of $(x, Q)$ points,
with the results averaged over multiple runs to suppress numerical fluctuations.

Fig.~\ref{fig:lhapdf-bench-performance} shows the average time as a function of the number of requested
points, up to $10^6$ evaluations. The top panel displays the absolute evaluation time whlile the bottom panel shows the ratio with respect to
\lhapdf. The \neopdf~results are shown for both the eager and lazy loading. While all interpolations scale linearly with the number of points, these
results clearly indicate that \neopdf~achieves a substantial performance improvement over \lhapdf. We see that \neopdf~(both lazy and eager mode)
reduces the average interpolation time by at least a factor of \underline{five}.

These results highlight one of the key advantages of the \neopdf~library: when using the exact interpolation method, it delivers the same numerical
accuracy as \lhapdf~but at significantly lower computational cost. Such a reduction directly translates into practical benefits for phenomenology
where PDF evaluations can dominate the runtime of Monte Carlo simulations.

\subsection{\tmdlib}
\label{subsec:benchmark-tmdlib}

We know benchmark \neopdf~against \tmdlib~for transverse momentum-dependent distributions. In contrast to the \lhapdf~benchmark in
the previous section, we always compare the built-in \tmdlib~interpolation---which is set
dependent---to the Cubic spline or Chebyshev implementation within \neopdf. The reason being that \neopdf~does not inherentlty
re-implement all the interpolation strategies supported by \tmdlib.

Given that \neopdf~provides an interface to \tmdlib~(see App.~\ref{app:cli}), all the
calls are performed via the \neopdf~methods. Furthermore, since \tmdlib~does not provide a suitable Python API, all of the
benchmarks are performed using the C++ interface, unless otherwise specified.

\paragraph{Interpolation accuracy.} In order to benchmark the relative accuracies of the Chebyshev interpolation in \neopdf~for TMD distributions,
we perform similar studies as the one presented in Sec.~\ref{sec:design}. Using the same representative functions
as in Eqs.\ref{eq:abmp} and~\ref{eq:herepdf20}, we account for
the intrinsic transverse momentum dependence using the \enquote{Gaussian Ansatz}~\cite{Bastami:2018xqd,Anselmino:2013lza,Anselmino:2011ch} in
which factorisation of $x$ and $k_T$ is assumed:
\begin{equation}
	x f_i (x, k_T) = x f_i (x) \frac{1}{\pi \langle k_T^2 (x) \rangle} \exp \left( - \frac{k_T^2}{\langle k_T^2 (x) \rangle} \right),
	\label{eq:guassian-ansatz}
\end{equation}
where $\langle k_T^2 (x) \rangle$ is the mean squared intrinsic transverse momentum---which may be treated as a constant or parametrised
with a mild $x$-dependence. In our study, we model its $x$-dependence as $\langle k_T^2 (x) \rangle = a + bx (1-x)$. By construction, the
Gaussian factor in Eq.~\ref{eq:guassian-ansatz} is normalised such that the integration over $k_T$ reproduces the collinear distribution
$\int 2 \pi k_T \mathrm{d} k_T f_i (x, k_T) =  f_i (x)$, ensuring consistency with the collinear limit.

The results are shown in Fig.~\ref{fig:tmdlib-interp-accuracy} for the two representative functions
$x f_1 (x, k_T)$ and $x f_2 (x, k_T)$ for $a=0.38$ and $b=0$. The top
panels show the results as a function of the longitudinal momentum fraction $x$ while the bottom panels show the results as a function of the
parton's intrinsic transverse momentum $k_T$. The exact functions that are being interpolated are shown by the black thick solid line with the
negative values represented by dashes.
The cubic spline interpolant is always expressed in terms of a single subgrid $\left[ 10^{-6}, 1 \right]_{(100)}$ with a total number of points
$n_{\rm pts} = 100$. The Chebyshev interpolant is instead expressed in terms of two subgrids for both $x$ and $k_T$ with $\left[ 10^{-6}, 0.2, 1 \right]_{(n,n)}$
and $\left[ 10^{-3}, 0.5, 2 \right]_{(n,n)}~\mathrm{GeV}$, respectively. Two grid configurations are considered, namely a low-density grid with $(25,25)$-point (purple)
and a high-density version with $(50,50)$-point (red), each containing a total number of grid points $n_{\rm pts} = 49$ and $n_{\rm pts} = 99$.
Similar conclusions to the one-dimensional case in Sec.~\ref{sec:design} can be drawn. Specifically, the Chebyshev interpolation
significantly outperforms the cubic Hermite spline by orders of magnitude, even with much smaller number of grid points.
 
\paragraph{\neopdf~benchmarks} To quantify the interpolation accuracy of \neopdf, we perform the benchmark using the MAP22 fragmentation
set~\cite{Bacchetta:2022awv} from the MAP collaboration. This is a set of unpolarised TMDFF distributions extracted from SIDIS data in which
large logarithms are resummed up to next-to-next-to-next-to-leading logarithmic (N$^3$LL) accuracy. Specifically, we use the negatively
charged Kaon set.

In order to perform the benchmarks, we convert the set into the \neopdf~format using the CLI (see App.~\ref{app:cli}). For the
purpose of the comparisons, the \neopdf~set is constructed using the following grid configurations:
\begin{align}
	x \:\: &: \left[ 10^{-1}, 0.2, 1 \right]_{(25, 25)}, \\
	k_T &: \left[ 10^{-4}, 0.5, 5\right]_{(25,25)}~\mathrm{GeV}, \\
	Q^2 &: \left[1.0, 4.2, 10^4 \right]_{(25,25)}~\mathrm{GeV}^2 ,
\end{align}
where we always resort to a two subgrid configuration for all of the kinematic variables.

\begin{figure}[!tb]
	\centering
	\includegraphics[width=0.495\textwidth]{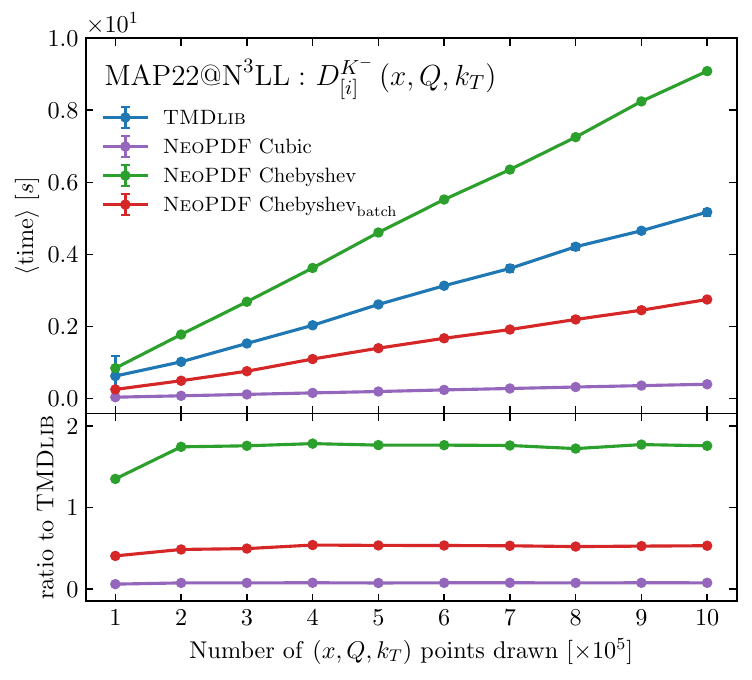}
	\caption{Benchmark of the interpolation evaluation time for the Kaon $K^-$ TMD distributions $D^{K^-}_{[i]}$ using the MAP22 fragmentation set
	at N$^3$LL. The time accounts for the computation of all the PDF flavours.
	Compared are the \tmdlib~results with the different interpolation methods in \neopdf. The average wall-clock time per evaluation is shown as a function
	of the number of $(x, Q, k_T)$ points (top) together with the ratio relative to \tmdlib~(bottom).}
	\label{fig:tmd-bench-performance}
\end{figure}

The benchmark results are shown in Fig.~\ref{fig:tmdlib-bench-accuracy} where the Kaon TMD distributions are shown for the strange (left)
and up (right) quarks. In both comparisons, the central panel shows the absolute values of $D^{K^-}_i(x, k_T)$ for the entire range of the parton's
intrinsic transverse momentum $k_T$ and longitudinal momentum fraction $x$. The left panel shows the relative difference between \neopdf's
Chebyshev interpolation and \tmdlib~as a function of $k_T$ for fixed values of $x$, namely $x=0.1$ and $x=0.2$. Similarly, the bottom panel shows
the relative difference as a function of $x$ for fixed values of $k_T$, namely $k_T=10^{-3}~\mathrm{GeV}$ and $k_T=2~\mathrm{GeV}$. All the results are computed for a fixed
hard scale $Q^2 = 10^4~\mathrm{GeV}^2$. We see that the relative differences in terms of interpolation accuracy between the two libraries---for
the considered TMDFF set---is generally found to be in the ballpark of $10^{-5}$ and $10^{-6}$. The reason why the relative interpolation accuracy
for $x=0.1$ reaches $10^{-7}$ is mainly because that specific point sits exactly on the grid knot.

\paragraph{Interpolation evaluation time.} In addition to accuracy, the efficiency of evaluating TMD functions is a critical requirement for practical
applications, especially given the three-dimensional nature of the interpolation. Fig.~\ref{fig:tmd-bench-performance} presents a benchmark of the
interpolation evaluation time for the MAP22 fragmentation set, comparing \neopdf~with \tmdlib. The average wall-clock time per evaluation is shown
as a function of the number of requested $(x, Q, k_T)$ points, with the ratios to \tmdlib~displayed in the bottom panel. The time accounts for the
computation of all the PDF flavours given that \tmdlib~always perform the interpolations with multiple flavours.

The \neopdf~results are shown for two interpolation strategies, namely the piecewise cubic spline and the global Chebyshev. The cubic method (orange)
is significantly faster than both the \neopdf~Chebyshev (green) and the \tmdlib~(blue) interpolations, owing to its local, piecewise nature in which
only the nearest knots are accessed to compute an interpolated value. This locality makes the cubic strategy highly efficient for multi-dimensional
interpolations, particularly when the evaluation involves a large number of independent points.

In contrast, the Chebyshev interpolation requires a polynomial reconstruction across the full phase-space domain for each evaluation. Consequently, its
evaluation time is higher than that of local cubic interpolation, and the overhead increases with the dimensionality of the interpolation space. Nonetheless,
the benchmarks demonstrate that NeoPDF’s Chebyshev implementation is only about two times slower than \tmdlib---which uses a cubic interpolation---across
the tested kinematic ranges.

However, a further advantage of the Chebyshev interpolation strategy is its suitability for batch optimisation. Owing to the global nature of the interpolation
coefficients, evaluations for multiple points can be computed in a vectorized manner, thereby minimising redundant calculations. Such an optimisation (red)
results in a substantial reduction in evaluation time while maintaining the characteristics of the Chebyshev interpolation.

These benchmark demonstrates that \neopdf~combines the efficiency of fast local cubic interpolation with more accurate global Chebyshev strategies,
thus providing a versatile and performant alternative for TMD distributions.


\section{Conclusions and outlook}
\label{sec:conclusions}

We have presented \neopdf, a modern interpolation framework for both collinear and transverse momentum–dependent PDFs with modern features,
designed to meet the precision and efficiency demands of current and future collider phenomenology. By unifying the treatment of diverse
non-perturbative inputs, \neopdf~enables consistent and unambiguous interpolations across different processes and kinematic domains.

We have demonstrated through benchmark studies that, when using the same piecewise cubic spline interpolations, \neopdf~reproduces
\lhapdf~results with agreement at the level of machine precision across the full $(x, Q^2)$--plane. This guarantees full backward compatibility
and seamless integration into existing workflows that use \lhapdf. Moreover, \neopdf~implements a Chebyshev-based global interpolation strategy,
which yields orders-of-magnitude improvements in accuracy with significantly fewer grid points. These gains translate directly into more compact
grids, and reduced storage requirements, while also providing numerically stable derivatives essential for various
precision QCD analyses.

For transverse momentum-dependent distributions, we have systematically benchmarked \neopdf~against \tmdlib. In particular, we show how the
Chebyshev interpolation is particularly suitable for TMD distributions, significantly outperforming standard piecewise methods by orders
of magnitude in accuracy, even with smaller grids. Furthermore, the benchmark results demonstrated that \neopdf~attains comparable or superior
evaluation time, yielding faster interpolation. These position \neopdf~as a reliable, versatile, and efficient alternative interpolation
library for TMD applications.

Future developments of \neopdf~will focus on extending its applicability to increasingly complex non-perturbative inputs
and the development of a centralised repository to upload and download \neopdf~sets.

The planned inclusion of GPDs and GTMDs will generalize the framework to fully multidimensional correlation functions, providing a unified tool for hadron tomography. The
implementation of analytical interpolations based on DGLAP evolution will reduce reliance on dense pre-tabulated grids by enabling the scale
dependence to be computed dynamically and consistently.
Furthermore, it would be interesting to explore the heterogeneous computing paradigms by exploiting the parallel processing capabilities of Graphics
Processing Units (GPUs)~\cite{Carrazza:2020qwu}.
Finally, the support for General-Mass Variable Flavour Number Schemes (GM-VFNS)~\cite{Kniehl:2004fy,Kniehl:2005mk}, such as
FONLL~\cite{Cacciari:1998it,Forte:2010ta,Barontini:2024xgu}, will ensure reliable and continuous treatments of heavy-flavour
thresholds.

Together, these extensions will consolidate NeoPDF as a scalable, precise, and future-proof infrastructure for QCD phenomenology
at the LHC, the EIC, and future colliders.

\paragraph{Acknowledgments:} The author wishes to acknowledge Valerio Bertone for inspiring the support for TMD distributions. The author gratefully
acknowledges Simone Rodini for various discussions regarding the design of \neopdf~and the Chebyshev interpolation. The author
also wishes to thank Juan M. Cruz-Martinez for stressing the relevance of the LHAPDF no-code migration. Finally, the author thanks Valerio Bertone,
Stefano Forte, Felix Hekhorn, Juan Rojo, and the members of the NNPDF collaboration for a careful reading of the manuscript and for providing
invaluable feedback. 
The work of T.R.R. was partially supported by an Accelerating Scientific Discoveries (ASDI2021) grant from the Netherlands eScience Center (NLeSC),
grant number 027.020.G05.

\appendix
\renewcommand\thesection{\Alph{section}} 
\renewcommand\thesubsection{\thesection.\arabic{subsection}} 
\renewcommand\thesubsubsection{\thesubsection.\arabic{subsubsection}} 

\section{Installing \neopdf}
\label{app:installation}

The following section provides detailed instructions for installing \neopdf~and its APIs for Rust, Python, C/C++, Fortran,
and Mathematica. Further details and instructions are available in the online code documentation
\href{https://qcdlab.github.io/neopdf/}{https://qcdlab.github.io/neopdf/}.

\subsection{Rust crate}

To use \neopdf~in a Rust project, simply add the following lines to the \texttt{Cargo.toml}:
\begin{lstlisting}[language=python]
[dependencies]
neopdf = "0.2.0"
\end{lstlisting}
Cargo will automatically fetch and compile the dependencies the next time the project is built with \texttt{cargo build}.

\subsection{Python API}

\neopdf~is available on the Python Package Index (PyPI) as \texttt{neopdf-hep} and therefore can be installed easily with any Python package
manager. For instance, using the \texttt{pip} package installer:
\begin{lstlisting}[language=bash, numbers=none]
> pip install neopdf-hep
\end{lstlisting}
Alternatively, the Python API can be built from source by cloning the repository and running the following command: 
\begin{lstlisting}[language=bash, numbers=none]
> maturin develop --manifest-path neopdf_pyapi/Cargo.toml --release --extras test
\end{lstlisting}
Note that this requires \texttt{maturin} to be installed in the environment.

\subsection{C/C++ API}

The simplest way to install the C/C++ API and the C++ Object Oriented Programming header is to download the pre-built libraries:
\begin{lstlisting}[language=bash, numbers=none]
> curl --proto '=https' --tlsv1.2 -sSf https://raw.githubusercontent.com/QCDLab/neopdf/refs/heads/master/install-capi.sh | sh
\end{lstlisting}
To pass the installation directory for where to put the files, change the arguments of the shell as follows:
\begin{lstlisting}[language=bash, numbers=none]
> .. | sh -s -- --prefix /custom/installation/path
\end{lstlisting}
By default, the script will download the latest stable release. Specific version can be installed by passing the version along with
the \texttt{--version} flag:
\begin{lstlisting}[language=bash, numbers=none]
> .. | sh -s -- --version 0.2.0
\end{lstlisting}

Alternatively, the C/C++ API can be built from source by first installing \texttt{cargo-c} by running the following command in the
root of the repository:
\begin{lstlisting}[language=bash, numbers=none]
> cargo install cargo-c
\end{lstlisting}
This will allow us to build and install C-ABI compatible dynamic and static libraries. Then, running the following command:
\begin{lstlisting}[language=bash, numbers=none]
> export CARGO_C_INSTALL_PREFIX=${prefix}
> cargo cinstall --manifest-path neopdf_capi/Cargo.toml --release --prefix=${prefix}
\end{lstlisting}
will produce and install a correct \texttt{pkg-config} file, a static and a dynamic library, and a C header in the \texttt{prefix}
path. This path can then be added to the \texttt{PKG\_CONFIG\_PATH} and \texttt{LD\_LIBRARY\_PATH} environment variables by running:
\begin{lstlisting}[language=bash, numbers=none]
> export LD_LIBRARY_PATH=${prefix}/lib:$LD_LIBRARY_PATH
> export PKG_CONFIG_PATH=${prefix}/lib/pkgconfig:$PKG_CONFIG_PATH
\end{lstlisting}

\subsection{Fortran API}

The Fortran API requires the C/C++ API to be installed. Once the C/C++ API is available, simply copy the \texttt{neopdf\_fapi/neopdf.f90}
into the working directory and generate the Fortran module with:
\begin{lstlisting}[language=bash, numbers=none]
> gfortran -c neopdf.f90
\end{lstlisting}
This will generate a \texttt{neopdf.mod} file which can be used in Fortran programs by including in the header:
\begin{lstlisting}[language=Fortran]
use neopdf
\end{lstlisting}

\subsection{Mathematica bindings}

\neopdf~provides bindings to the Wolfram LibraryLink interface, making it possible to call \neopdf~'s Rust APIs from the Wolfram
Mathematica language. To install the Wolfram bindings, simply run the following command:
\begin{lstlisting}[language=bash, numbers=none]
> cargo build --release --manifest-path neopdf_wolfram/Cargo.toml
\end{lstlisting}
Note that this requires Wolfram Mathematica to be installed in the system. Examples of how the bindings are used are shown in
Sec.~\ref{app:cli}.

\subsection{CLI interface}

The command line interface (CLI) to the \neopdf~'s API is also available on PyPI as \texttt{neopdf-cli}. It can therefore be easily
installed as:
\begin{lstlisting}[language=bash, numbers=none]
> pip install neopdf-cli
\end{lstlisting}
Similar to then other packages, the CLI tool can be built from source for development purposes. To build and install the \neopdf~CLI
interface, simply run:
\begin{lstlisting}[language=bash, numbers=none]
> cargo install --path neopdf_cli
\end{lstlisting}
This will compile the CLI in debug mode and make the \texttt{neopdf} command available in the cargo bin directory (usually in
\texttt{\textasciitilde/.cargo/bin}).

\subsection{\neopdf~data path}

By default, \neopdf~stores and loads PDF sets from \texttt{\$HOME/.local/share/neopdf}. This can however be overwritten via the
environment variable \texttt{NEOPDF\_DATA\_PATH} to point to the \lhapdf~path for example, or any other path:
\begin{lstlisting}[language=bash, numbers=none]
> export NEOPDF_DATA_PATH=${LHAPDF_DATA_PATH}
\end{lstlisting}

\section{Example usage of the CLI}
\label{app:cli}

The \neopdf~CLI tool provides a command-line interface to the \neopdf~library, enabling users to perform PDF interpolation, $\alpha_s$
evaluation, file format conversion, and metadata inspection directly from the terminal. The following section provides examples of the
standard usage of the CLI. For more advanced usage, refer to the helper by running the command \texttt{neopdf --help}.

\begin{figure*}[tb]
	\centering
	\includegraphics[width=0.495\textwidth]{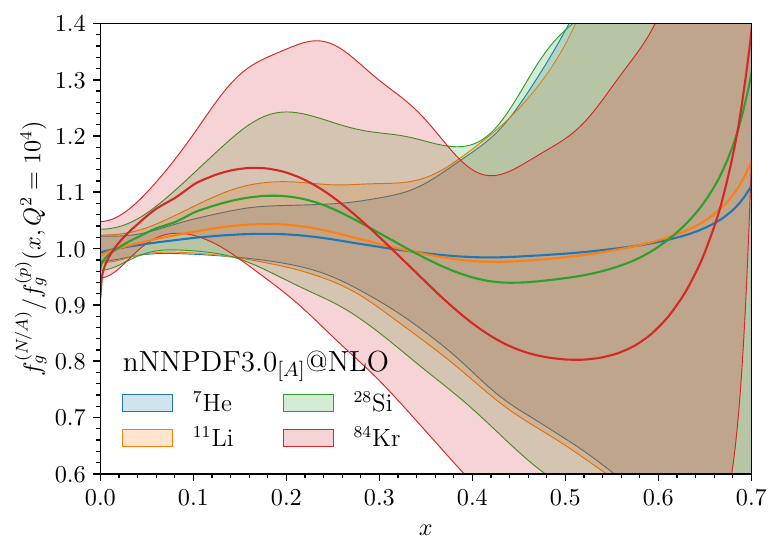}
	\includegraphics[width=0.495\textwidth]{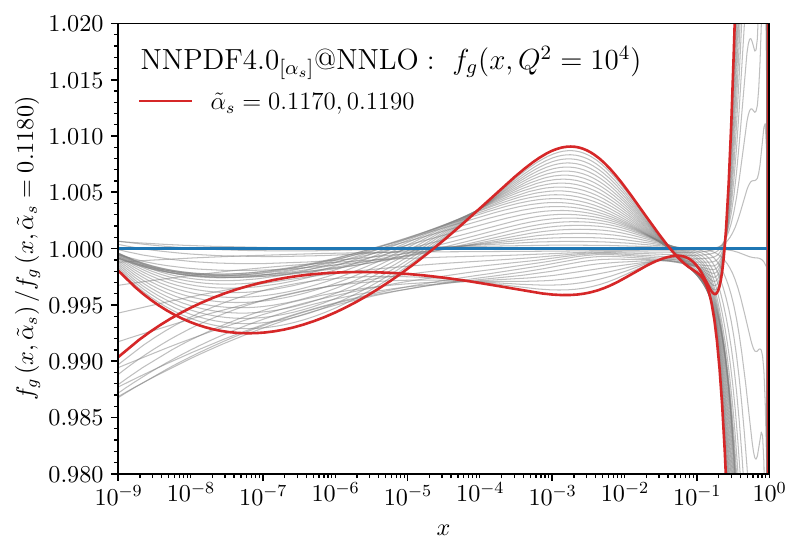}
	\caption{Examples of usage of the combined nuclear (left) and reference strong coupling sets. The sets are the combination of
	all the available nuclear PDF and $\tilde{\alpha}_s \equiv \alpha_s(M_Z)$ variation sets, using nNNPDF3.0 and NNPDF4.0
	respectively. The left plot shows the interpolated nPDFs for values of nuclei for which predictions are not available, while
	the right plot shows the interpolated results for $n=40$ $\tilde{\alpha}_s$ values within the range $[0.1170, 0.1190]$.}
	\label{fig:combined-lhapdfs}
\end{figure*}

\subsection{Standard usage}

One of the main usages of the CLI is to inspect a set's metadata information. Using the \textsc{NNPDF4.0} set from LHAPDF:
\begin{lstlisting}[language=bash, numbers=none]
> neopdf read metadata NNPDF40_nnlo_as_01180

Set Description: NNPDF4.0 NNLO global fit, ...
Set Index: 331100
Number of Members: 101
XMin: 0.000000001
XMax: 1
QMin: 1.65
QMax: 100000
Flavors: [-5, -4, -3, -2, -1, 21, 1, 2, 3, 4, 5]
Format: lhagrid1
...
Polarized: false
Set Type: SpaceLike
Interpolator Type: LogBicubic
...
\end{lstlisting}
where for visual purposes we have omitted some of the printed out information. We can further check how many subgrids does the set contain:
\begin{lstlisting}[language=bash, numbers=none]
> neopdf read num_subgrids NNPDF40_nnlo_as_01180

  2
\end{lstlisting}
We can now inspect the contents of the first subgrid for the central member of the gluon PDF:
\begin{lstlisting}[language=bash, numbers=none]
> neopdf read subgrid NNPDF40_nnlo_as_01180 --member=0 --subgrid-index=0 --pid=21

  [x | Q2]   2.72250e0   3.19494e0   3.77488e0  ...
1.00000e-9  1.48441e-1  -1.47266e0  -3.42816e0  ...
1.29708e-9  1.53954e-1  -1.36579e0  -3.16487e0  ...
1.68243e-9  1.59670e-1  -1.26122e0  -2.90961e0  ...
...
 1.00000e0   0.00000e0   0.00000e0   0.00000e0  ...
\end{lstlisting}
One of the prime features of the \neopdf~CLI is the ability to directly compute the interpolated function $x f(x, Q^2)$ for a given
member and PDF flavour:
\begin{lstlisting}[language=bash, numbers=none]
> neopdf compute xfx_q2 NNPDF40_nnlo_as_01180 --member=0 --pid=21 1e-3 10.0

  7.1276606679158565
\end{lstlisting}
where the last two arguments are simply the $x$ and $Q^2$ values at which the PDF should be evaluated.

The \neopdf~CLI tool can convert existing PDF sets in the LHAPDF format into the \neopdf's format to leverage the efficient compression
and data access. This can be done by running the following command:
\begin{lstlisting}[language=bash, numbers=none]
> neopdf write convert NNPDF40_nnlo_as_01180 --output NNPDF40_nnlo_as_01180.neopdf.lz4
\end{lstlisting}
This will generate the a \texttt{.neopdf.lz4} file in the current directory which can then be moved into the \neopdf~data path. We note
that all of the previous commands work with \neopdf~grids by simply appending \texttt{.neopdf.lz4} to the PDF names.

\subsection{Combining LHAPDF sets}

\neopdf~provides a CLI interface to combine multiple LHAPDF sets---which differ in values of $A$ (respectively $\tilde{\alpha}_s)$---into a single
\neopdf~grid in order to explicitly account for the $A$-dependence (respectively $\tilde{\alpha}_s$-dependence). For instance, to combine sets
with $\tilde{\alpha}_s$ variations, we first need to write a file (say \texttt{alphas-sets.txt}) that contains the name of the PDF sets:
\begin{lstlisting}[language=bash]
NNPDF40_nnlo_as_01160
NNPDF40_nnlo_as_01170
NNPDF40_nnlo_as_01175
NNPDF40_nnlo_as_01180
NNPDF40_nnlo_as_01185
NNPDF40_nnlo_as_01190
\end{lstlisting}
We can then pass the file as input to the CLI in order to combine them into one single set:
\begin{lstlisting}[language=bash, numbers=none]
> neopdf write combine-alphas --names-file alphas-sets.txt --output NNPDF40_nnlo_as_combined.neopdf.lz4
\end{lstlisting}
The same syntactic structure can be used combine multiple nuclear PDF sets but using \texttt{combine-npdfs}. Fig.~\ref{fig:combined-lhapdfs}
illustrates examples of usage of the combine sets. The left plot shows nuclear PDF predictions for values of $A$ which are not prodived by
\textsc{nNNPDF3.0} while the right plots shows more $\tilde{\alpha}_s$ variations within the range $\left[ 0.01170, 0.1190 \right]$.

\subsection{Interface to \tmdlib}
\label{app:tmdlib-interface}

\neopdf~provides an interface to the \tmdlib~library which allows users for example to interpolate TMD PDFs from the CLI. Most importantly,
it can be used to easily convert \neopdf~sets into \neopdf~formats. To install the \tmdlib~interface, we use the same command to install
the CLI but turning on the \tmdlib~feature:
\begin{lstlisting}[language=bash, numbers=none]
> cargo install --path neopdf_cli --features=tmdlib
\end{lstlisting}
Note that this requires \tmdlib~and its dependencies to be installed and available in the environment paths. We can now  interpolate
\tmdlib~sets from the CLI, the command is similar to the one used for regular sets:
\begin{lstlisting}[language=bash, numbers=none]
> neopdf compute xfx_q2_kt MAP22_grids_FF_Km_N3LL --member=0 --pid=2 1e-3 1e-1 4e3

  0.01477369348062961
\end{lstlisting}
where the last three arguments represent the parton's intrinsic transverse momentum $k_T$, the momentum fraction $x$, and the hard scale
$Q^2$.

To convert a \tmdlib~set into the \neopdf~format, one needs to write a configuration file in the \texttt{.toml} format. This configuration
file should contain all the information about the metadata and the grid configurations. The following is an example using the same TMD set
(\texttt{MAP22.toml}):
\begin{lstlisting}[language=Python]
set_name = "MAP22_grids_FF_Km_N3LL"
set_desc = "NeoPDF TMDFFs for K-"
set_index = 0
n_members = 2

# Inner edges for the subgrids.
x_inner_edges = [0.2]
q_inner_edges = [] # Q, not Q2
kt_inner_edges = [1e-2, 1.0]

# Number of points for (subg)grids.
n_x = [5, 5]
n_q = [6]
n_kt = [5, 5, 4]

# Axes that are not part of the TMD interpolation
nucleons = [0.0] # dummy value
alphas = [0.118] # alpha_s(M_Z)

# Metadata
pids = [-3, -2, -1, 21, 1, 2, 3]
polarised = false
set_type = "TimeLike"
interpolator_type = "LogChebyshev" # or LogTricubic
error_type = "replicas"
hadron_pid = 321 # Kaon
alphas_qs = [91.1876] # dummy values
alphas_vals = [0.118] # dummy values

# Physics Parameters
flavor_scheme = "fixed"
order_qcd = 2
alphas_order_qcd = 2
m_w = 80.352
m_z = 91.1876
m_up = 0.0
m_down = 0.0
m_strange = 0.0
m_charm = 1.51
m_bottom = 4.92
m_top = 172.5
alphas_type = "ipol"
number_flavors = 4
\end{lstlisting}
The set can now be converted by running the following command:
\begin{lstlisting}[language=bash, numbers=none]
> neopdf write convert-tmd --input MAP22.toml --output MAP22_grids_FF_Km_N3LL.neopdf.lz4
\end{lstlisting}
Similar to standard PDF sets, once the conversion has been performed successfully (and the converted \neopdf~set available in the
\texttt{NEOPDF\_DATA\_PATH}), we can inspect the contents of the first subgrid for the first $k_T$ knot to make sure that everything
is correct:
\begin{lstlisting}[language=bash, numbers=none]
> neopdf read subgrid MAP22_grids_FF_Km_N3LL.neopdf.lz4 --member=0 --subgrid-index=0 --pid=2 --kt-index=0

Displaying grid for kT = 0.0001

  [x | Q2]   1.00000e0   2.40972e0   2.40972e1 ...
1.00000e-1 -6.70867e-1 -4.80573e-1 -1.53370e-1 ...
1.10684e-1 -7.26434e-1 -4.81932e-1 -1.40332e-1 ...
1.41421e-1 -7.37791e-1 -3.84220e-1 -8.14145e-2 ...
1.80695e-1 -5.23889e-1 -1.84548e-1 -1.28827e-2 ...
2.00000e-1 -3.89947e-1 -1.06392e-1  7.02618e-3 ...
\end{lstlisting}

\section{API usage examples}

The following section provides some examples on how to use the various APIs to interpolate and perform some operations on PDF sets.
The current examples only provide the most basic usage, and in particular mainly showcases the compatibility layers with \lhapdf~for
no-code migration. For more advanced usage and additional features, refer to the
documentation\footnote{\faBook~Documentation: \href{https://qcdlab.github.io/neopdf/}{https://qcdlab.github.io/neopdf/}}.

\subsection{Python API}

The following Python examples illustrate the no-code migration when transitionning from \lhapdf~while giving a brief overview of
some of the additional features, such as converting a \lhapdf~set into the \neopdf~format.

\begin{lstlisting}[language=Python]
from neopdf.pdf import PDF as lhapdf

# To load NeoPDF sets, append with `.neopdf.lz4`
# Use `mkPDFs` to load all the members
pdf = lhapdf.mkPDF("NNPDF40_nnlo_as_01180")

# Get the Metadata of the PDF set as a dictionary
metadata = pdf.metadata().to_dict()
print(metadata.keys())

# Print the shape of the first subgrid
pdf_subgrids = pdf.subgrids()
print(pdf_subgrids[0].grid_shape())

# Compute the interpolated value `xf(x,Q2)`
xf = pdf.xfxQ2(21, 1e-5, 1e3)
# Compute the value of strong coupling
a_s = pdf.alphasQ2(10)
print(xf, a_s)

# Convert LHAPDF set into NeoPDF format
from neopdf.converter import convert_lhapdf

convert_lhapdf(
 pdf_name="NNPDF40_nnlo_as_01180",
 output_path="NNPDF40_nnlo_as_01180.neopdf.lz4"
)

# After moving the set into `NEOPDF_DATA_PATH`
# we can load and check the new NeoPDF set
neo_pdf = lhapdf.mkPDF("NNPDF40_nnlo_as_01180.neopdf.lz4")
xf_neo = neo_pdf.xfxQ2(21, 1e-5, 1e3)
print(xf_neo)
\end{lstlisting}
\neopdf~also provides a function, \texttt{xfxQ2s()}, that takes as inputs lists of PID, $x$, and $Q^2$ values and
computes the interpolated values using their Cartesian products.

In presence of TMD sets, or distributions that depend on more parameters than just the longitudinal momentum fraction $x$
and hard scale $Q^2$, the difference lies in how the interpolation function is called.
\begin{lstlisting}[language=Python]
from neopdf.pdf import PDF as tmdlib

pdf = tmdlib.mkPDF("MAP22_grids_FF_Km_N3LL.neopdf.lz4")

# Compute the interpolated value `xf(kT,x,Q2)`
xf_kt = pdf.xfxQ2_ND(21, params=[0.5, 1e-5, 1e3])
print(xf_kt)
\end{lstlisting}
where \texttt{params} can take a list of varying length depending on the variables that the distribution depends on
provided that the order \texttt{[A,alphas,kT,x,Q2]} is preserved. For instance, for a nuclear TMD set, the \texttt{params}
argument takes the following values: \texttt{[A,kT,x,Q2]}.

\subsection{C/C++ API}

\begin{lstlisting}[language=C++]
#include <NeoPDF.hpp>
#include <iostream>

using namespace neopdf;
using namespace NEOLHAPDF; // LHAPDF compatibility

// To load NeoPDF sets, append with `.neopdf.lz4`
std::string PDFNAME = "NNPDF40_nnlo_as_01180";

// Pure C LHAPDF compatibility
void neopdf_c_lhapdf_compatibility() {
  initpdfsetbyname(PDFNAME.c_str());
  initpdf(0); // init. PDF member

  double xfs[13];
  evolvepdf(1e-3,  31.623, xfs);
  double a_s = alphaspdf(3.163);
  std::cout << "xf=" << xfs[0] << " ; a_s=" << a_s << std::endl;
}

// Pure C native NeoPDF syntax
void neopdf_native_c() {
  NeoPDFWrapper* pdf = neopdf_pdf_load(PDFNAME.c_str(), 0);

  double xf = neopdf_pdf_xfxq2(pdf, 21, 1e-5, 1e3);
  double a_s = neopdf_pdf_alphas_q2(pdf, 10);
  std::cout << "xf=" << xf << " ; a_s=" << a_s << std::endl;
  
  // free object from memory
  neopdf_pdf_free(pdf);
}

// C++ OOP with LHAPDF compatibility
void neopdf_cpp_oop_lhapdf_compatibility() {
  PDF* pdf = mkPDF(PDFNAME, 0);

  // Use `xfxQ2s` for input vectors
  double xf = pdf->xfxQ2(21, 1e-5, 1e3);
  double a_s = pdf->alphasQ2(10);
  std::cout << "xf=" << xf << " ; a_s=" << a_s << std::endl;
  
  // free object from memory
  delete pdf;
}

// C++ OOP with native NeoPDF synatex
void neopdf_native_cpp_oop() {
  NeoPDF pdf("NNPDF40_nnlo_as_01180", 0);

  // Use `xfxQ2s` for input vectors
  double xf = pdf.xfxQ2(21, 1e-5, 1e3);
  double a_s = pdf.alphasQ2(10);
  std::cout << "xf=" << xf << " ; a_s=" << a_s << std::endl;
}

int main() {
  neopdf_c_lhapdf_compatibility();
  neopdf_native_c();
  neopdf_cpp_oop_lhapdf_compatibility();
  neopdf_native_cpp_oop();

  return 0;
}
\end{lstlisting}
Similar to the Python API, the C/C++ API provides a \texttt{pdf.xfxQ2\_ND()} function to interpolate distributions
that depend on more parameters than just the longitudinal momentum fraction $x$ and the hard scale $Q^2$.

\subsection{Fortran API}

The following Fortran example illustrates a very basic usage of \neopdf~that showcases the \lhapdf~drop-in parity.
For more complex usage such as interpolating TMD distributions or filling and writing a \neopdf~grid, refer to the
code documentation.

\begin{lstlisting}[language=Fortran]
program check_fapi
  use neopdf
  use iso_c_binding
  
  implicit none
  
  character(len=256) :: pdf_name
  double precision   :: x, xfg, a_s
  double precision xfs(-6:6)
  
  pdf_name = "NNPDF40_nnlo_as_01180"
  
  call initpdfsetbyname(pdf_name)
  call initpdf(0) ! Init. central member
  call evolvepdf(x, sqrt(1e3), xfs)

  xfs = xfs(0)
  a_s = alphaspdf(sqrt(10.0))
  write(*,'(ES10.3, 2X, ES12.5, 2X, ES12.5)') x, xfg, a_s
  
  call neopdf_pdf_free(pdf)
end program check_fapi
\end{lstlisting}

\subsection{Mathematica interface}

\neopdf~provides a very basic interface to the Wolfram language allowing the Rust APIs to be used in Mathematica.
\begin{lstlisting}[language=Mathematica]
libPath="/path/to/libneopdf_wolfram.dylib";

loadPDF = LibraryFunctionLoad[
  libPath,
  "NeoPDF_Load",
  {"UTF8String", Integer},
  Integer
];
xfxq2 = LibraryFunctionLoad[
  libPath,
  "NeoPDF_XFXQ2",
  {Integer, Integer, {Real, 1}},
  Real
];
alphasQ2 = LibraryFunctionLoad[
  libPath,
  "NeoPDF_AlphasQ2",
  {Integer, Real},
  Real
];
clearPDFs = LibraryFunctionLoad[
  libPath,
  "NeoPDF_Clear",
  {},
  "Void"
];

pdfName = "NNPDF40_nnlo_as_01180";
pdfIndex = loadPDF[pdfName, 0];

params = {0.01, 100.0};
resultXFXQ2 = xfxq2[pdfIndex, 21, params];
Print["xfxq2=", resultXFXQ2];

resultAlphaS = alphasQ2[pdfIndex, 100.0];
Print["alpha_s=", resultAlphaS];

clearPDFs[];
\end{lstlisting}

\section{Writing \neopdf~Grids}

The following section provides examples on how to write \neopdf~grids using the Python and C++ APIs. For more
details and examples with different programming languages bindings, refer to the tutorials section in the code
documentation. These include writing \neopdf~grids with complex subgrids structure or grids that depend on more
parameters than just the longitudinal momentumf fraction $x$ and the hard scale $Q^2$.

\subsection{Python API}

\begin{lstlisting}[language=Python]
from neopdf.gridpdf import GridArray, SubGrid
from neopdf.metadata import InterpolatorType, SetType, MetaData, PhysicsParameters
from neopdf.writer import compress

# Construct the shared Metadata
num_members = 10
x_min = 1e-5
x_max = 1.0
q_min = 4.0
q_max = 1e5
flavors = [nf for nf in range(-4, 5)]
alphas_vals = np.random.uniform(0.1, 0.2, 6)
alphas_qvalues = np.geomspace(q_min, q_max, 6)


# Define the Physical Parameters in the Metadata
physparams_kwargs = {
  "flavor_scheme": "fixed",
  "order_qcd": 2,
  "alphas_order_qcd": 2,
  "m_w": 80.3520,
  "m_z": 91.1876,
  "m_up": 0.0,
  "m_down": 0.0,
  "m_strange": 0.0,
  "m_charm": 1.51,
  "m_bottom": 4.92,
  "m_top": 172.5,
}
physparams = PhysicsParameters(**physparams_kwargs)

# Construct the PDF Metadata
metadata_kwargs = {
  "set_desc": "Toy NeoPDF set",
  "set_index": 123456,
  "num_members": num_members,
  "x_min": x_min,
  "x_max": x_max,
  "q_min": q_min,
  "q_max": q_max,
  "flavors": flavors,
  "format": "neopdf",
  "alphas_q_values": alphas_qvalues,
  "alphas_vals": alphas_vals,
  "polarised": False,
  "set_type": SetType.SpaceLike,
  "interpolator_type": InterpolatorType.LogBicubic,
  "phys_params": physparams,
}
metadata = MetaData(**metadata_kwargs)

# Construct the PDF Grid:
nucleons = [1]      # Proton Only
alphas_mZ = [0.118] # Only one value
kts = [0.0]         # No dependence on kT
grid_members = []

x_values = np.geomspace(x_min, x_max, 50)
# Subdivide the Q2 range into subgrids
q_mid = int(q_max / 3)
q2_sub1 = np.geomspace(q_min**2, q_mid**2, 25)
q2_sub2 = np.geomspace(q_mid**2, q_max**2, 25)
q2_values = [q2_sub1, q2_sub2]

for _ in range(num_members):
  sub_grids = []
  for q2_subgrid in q2_values:
    grid_shape = (
        len(nucleons),
        len(alphas_mZ),
        len(flavors),
        len(kts),
        x_values.size,
        q2_subgrid.size,
    )
    # simulate grid values with random numbers
    grid = np.random.uniform(0, 1, grid_shape)

    sub_grid = SubGrid(
        xs=x_values,
        q2s=q2_subgrid,
        kts=kts,
        nucleons=nucleons,
        alphas=alphas_mZ,
        grid=grid,
    )
    sub_grids.append(sub_grid)

  grid_member = GridArray(pids=flavors, subgrids=sub_grids)
  grid_members.append(grid_member)

# Write the compressed PDF set into disk
compress(grids=grid_members, metadata=metadata, path="TOY_NEOPDF.neopdf.lz4")
\end{lstlisting}
In the above examples, we have simulated the grid values with some random numbers while in
practice one needs to fill the subgrid for a given flavour, $x$, and $Q^2$. The order of
the loops follows the LHAPDF convention, i.e. \texttt{[x][Q2][pid]} with \texttt{x} being
the most outer loop and \texttt{pid} the most inner.

\subsection{C++ API}

In the following example, we show how to write \neopdf~grids in C++ using the Chebyshev
subgrids structure and interpolation.

\begin{lstlisting}[language=C++]
#include <NeoPDF.hpp>
#include <cassert>
#include <cmath>
#include <cstddef>
#include <cstdint>
#include <cstdio>
#include <cstdlib>
#include <cstring>
#include <iostream>
#include <random>
#include <tuple>
#include <vector>

using namespace neopdf;

std::vector<double> create_logchebyshev_grid(int n_points, double min, double max) {
 double u_min = std::log(min);
 double u_max = std::log(max);

 std::vector<double> grid_points;
 grid_points.reserve(n_points);

 for (int j = 0; j < n_points; ++j) {
  double dn =  static_cast<double>(n_points-1);
  double t_j = std::cos(M_PI*(n_points-1-j)/dn);
  double u_j = u_min+(u_max-u_min)*(t_j+1.0)/2.0;
  grid_points.push_back(std::exp(u_j));
 }
 
 return grid_points;
}

int main() {
 std::random_device rd;
 std::mt19937 gen_rndng(rd());
 std::uniform_real_distribution<double> rndng(0.0, 1.0);

 // Create an instance of grid writer
 GridWriter writer;

 // Set some parameters
 const std::size_t nb_members = 10;
 std::vector<int32_t> pids = {-6, -5, -4, -3, -2, -1, 0, 1, 2, 3, 4, 5, 6};
 double xmin = 1e-9, xmax = 1.0;
 double q2min = 1.0, q2max = 1e3;

 std::vector<std::tuple<double, double>> x_subgrids = {
  std::make_tuple(xmin, 0.2),
  std::make_tuple(0.2, xmax)        
 };
 std::vector<std::tuple<double, double>> q2_subgrids = {
  std::make_tuple(q2min, 4.92),
  std::make_tuple(4.92, q2max)        
 };

 for (size_t m = 0; m < nb_members; ++m) {
  // Start a new grid for the current member
  writer.new_grid();

  // Loop over the subgrids
  for (const auto& x_sub : x_subgrids){
   for (const auto& q2_sub: q2_subgrids) {
    auto xs = create_logchebyshev_grid(
     25, // number of subgrid points
     std::get<0>(x_sub),
     std::get<1>(x_sub)
    );
    auto q2s = create_logchebyshev_grid(
     25, // number of subgrid points
     std::get<0>(q2_sub),
     std::get<1>(q2_sub)
    );

    std::vector<double> grid_data;
    for (double x : xs) {
     for (double q2 : q2s) {
      for (int pid : pids) {
       double _rndm = rndng(gen_rndng);
       grid_data.push_back(abs(pid)*x*q2*_rndm);
      }
     }
    }

    // Add to the subgrid
    writer.add_subgrid(
     {1.0},    // A
     {0.1180}, // alphas
     {0.0},    // kT
     xs,
     q2s,
     grid_data
    );
   }
  }

  // Finalize the Grid for this member.
  writer.push_grid(pids);
 }

 std::vector<double> alphas_qs = {91.1876};
 std::vector<double> alphas_vals = {0.118};

 PhysicsParameters phys_params = {
  .flavor_scheme = "fixed",
  .order_qcd = 2,
  .alphas_order_qcd = 2,
  .m_w = 80.352,
  .m_z = 91.1876,
  .m_up = 0.0,
  .m_down = 0.0,
  .m_strange = 0.0,
  .m_charm = 1.51,
  .m_bottom = 4.92,
  .m_top = 172.5,
  .alphas_type = "ipol",
  .number_flavors = 4,
 };

 MetaData meta = {
  .set_desc = "Toy NeoPDF set",
  .set_index = 1234,
  .num_members = nb_members,
  .x_min = xmin,
  .x_max = xmax,
  .q_min = sqrt(q2min),
  .q_max = sqrt(q2max),
  .flavors = pids,
  .format = "neopdf",
  .alphas_q_values = alphas_qs,
  .alphas_vals = alphas_vals,
  .polarised = false,
  .set_type = NEOPDF_SET_TYPE_SPACE_LIKE,
  .interpolator_type = NEOPDF_INTERPOLATOR_TYPE_LOG_CHEBYSHEV,
  .error_type = "replicas",
  .hadron_pid = 2212,
  .phys_params = phys_params,
 };

 // Check if `NEOPDF_DATA_PATH` is defined and store the Grid there.
 const char* filename = "TOY_PDF.neopdf.lz4";
 const char* neopdf_path = std::getenv("NEOPDF_DATA_PATH");
 std::string output_path = neopdf_path
  ? std::string(neopdf_path) + (std::string(neopdf_path).back() == '/' ? "" : "/") + filename
  : filename;

 // Write the PDF Grid into disk
 try {
  writer.compress(meta, output_path);
  std::cout << "Compression succeeded!\n";
 } catch (const std::runtime_error& err) {
  std::cerr << "Compression failed: " << err.what() << "\n";
  return EXIT_FAILURE;
 }

 return EXIT_SUCCESS;
}
\end{lstlisting}

\bibliography{draft.bib}


\end{document}